\documentclass[12pt]{iopart}
\usepackage{iopams}
\usepackage{graphicx}
\usepackage{braket}
\usepackage{float}
\expandafter\let\csname equation*\endcsname\relax
\expandafter\let\csname endequation*\endcsname\relax
\usepackage{amsmath}
\usepackage{xcolor}
\usepackage{verbatim}
\usepackage{subfigure}
\usepackage{cite}
\usepackage{hyperref} 
\PassOptionsToPackage{pagebackref=true}{hyperref}
\hypersetup{
     colorlinks=true,
     citecolor=blue,
     urlcolor=orange,
     linkcolor=red
}
\usepackage{mycommand}
\usepackage{tikz}
\usepackage{circuitikz}
\usetikzlibrary{arrows,decorations.markings}
\usetikzlibrary{positioning}

\usepackage{pstricks}
\newrgbcolor{dred}{0.8 0.1 0.0}
\def\ra{$\rightarrow$}

\def\i{\item}

\def\simp{\text{simp}}
\def\eff{\text{eff}}
\def\ratio{\text{ratio}}
\def\net{\text{net}}

\def\site{\text{site}}
\def\Reff{R_\text{eff}}
\def\Geff{G_\text{eff}}

\def\rect{\text{rect}}
\def\hex{\text{hex}}

\renewcommand{\figurename}{Fig.} 
\def\FIGDIR{.}
\begin{document}
%
%

\title{Effective resistances of two dimensional resistor networks}
\author{Rajat Chandra Mishra{$^1$} and Himadri Barman{$^2$}}

\address{$^1$ St Stephens College, University of Delhi, New Delhi 110007, India}
\address{$^2$Department of Physics, Zhejiang University, Hangzhou 310027, China}
\ead{\mailto{hbarhbar@gmail.com}}
\date{\today}
%
\begin{abstract}
We investigate the behavior of two dimensional resistor networks, with finite sizes and different kinds (rectangular, hexagonal, and triangular) of lattice geometry. We construct the network by having a network-element repeat itself $L_x$ times in $x$-direction and $L_y$ times in the $y$-direction. We study the relationship between the effective resistance ($\Reff$) of the network on dimensions $L_x$ and $L_y$. The behavior is simple and intuitive for a network with rectangular geometry, however, it becomes non-trivial for other geometries which are solved numerically. We find that $\Reff$ depends on the ratio $L_x/L_y$ in all the three studied networks. We also check the consistency of our numerical results experimentally for small network sizes. 
\end{abstract}
%

\submitto{\EJP}

\maketitle
\section{Introduction}
Resistor network problems have been widely studied in various contexts, starting from textbook physics and competitive tests~\cite{book:irodov81} to  electrical engineering~\cite{book:bird10}, condensed matter physics~\cite{lai:etal:sci10}, and statistical physics~\cite{wu:etal:pre05}. Since many regular electrical networks take shapes of meshes, similar to the lattices in solid state crystals, it is intriguing to find out the equivalent or effective resistance of such networks. There had been extensive studies on two-dimensional lattices in order to investigate percolation based conductivity~\cite{kirkpatrick:rmp73} in such systems and methods like effective medium theory~\cite{kirkpatrick:rmp73,bernasconi:prb74,koplik:jpcssp81,toledo:etal:ces92} and Green's function method~\cite{kirkpatrick:rmp73,wu:bradley:pre94,cserti:ajp00,owaidat:asad:citp19} have been formulated. However, most of these studies focus on the infinite systems with stochastic resistance distribution (random resistor network). Although a few studies have been conducted recently for finite size networks, such studies either investigated equivalent resistance between two points inside the network or for networks that do not obey the typical crystal lattice symmetries~\cite{wu:jpamag04,tan:etal:jpamat13,essam:etal:rsos15,owaidat:asad:citp19,tan:tan:citp20,tan:tan:ps20}. Hence dimensional dependence of effective resistance ($\Reff$) for various geometries deserves separate attention and it also bears an academic interest to show how 
$\Reff$ can simply be estimated by solving a set of electrical equations.  
Network geometry dependence of $\Reff$ brings the connection to the graph theory~\cite{kagan:ajp15,dorfler:etal:ieee18} and generalization of $Y$-$\Delta$ or star-polygon transformation can open doors of future research~\cite{book:bollobas13}. The networks described in our paper are very straightforward and easy to solve numerically once the correct equations are formulated. However, such solutions do not exist in the literature to the best of our knowledge and hence our findings are both pedagogical and research oriented. Given resources, the network models can be constructed by students easily and knowing the dependence of $\Reff$ on the geometry, a device can be designed whose resistance can be controlled by tuning its dimensions. 

Our paper is organized in the following way. We first explain the generic resistor network setup and then discuss the analytical solution for the rectangular geometry. Then we discuss the numerical formulation and $\Reff$'s dependence on the dimensions, obtained from our numerical results for rectangular, hexagonal, and triangular resistor networks. As a summary, we compare these results for these three different geometries and finally we describe a small experiment to test our theoretical findings.

\section{Generic resistor network configuration}
\label{sec:gen:config}
We describe below our generic setup for various lattice geometries:
\begin{enumerate}
\i Define a two dimensional (2D) lattice. Though the geometry varies lattice to lattice, we define the size of each lattice by two Cartesian lengths $L_x$ and $L_y$. For a rectangular lattice, $L_x L_y$ becomes the number of the lattice points or sites as well.
\i Each lattice point attaches to a resistance of value $R$ spreading along the direction of its neighborhood lattice points.
\i  Apply a bias $V$ at one edge of the lattice (say in the direction of the length $L_x$) and ground the other edge (hence the lattice acts like an active medium attached to a battery or applied voltage). Only one point from each unit cell of the lattice is attached to the bias or grounding and such points must be equivalent for all unit cells of the lattice. 
\end{enumerate}

Typically, electrical networks with resistors and biases are solved by using one of the Kirchhoff's 
laws of circuit (originally announced by Gustav Kirchhoff in 1845)~\cite{kirchhoff:adp1847,kirchhoff:trans58}, which is often termed as mesh-current or nodal analysis by electrical engineers~\cite{book:bird10}. Out of Kirchhoff's voltage and current laws, it is more convenient to use the current law that states that total current at a circuit junction (lattice point or site in our case) must be zero. Hence the key equation for the Kirchhoff's current law (KCL) at a site or node $(i,j)$ in 2D Cartesian coordinate:
\blgn
\sum_k I_{ij}(k)=0 \Ra \sum_k [(V_{ik}-V_{ij})/R_{ik}+(V_{kj}-V_{ij})/R_{jk}]=0\,
\elgn
where $k$ denotes all nearest neighbor nodes (lattice points, voltage or grounding connection) to the site $(i,j)$. We discuss the implementation of this in the forthcoming sections, where we formulate them in the form of a matrix equation for various network geometries.

\section{Rectangular resistor network}
\label{sec:rect:network}
\begin{figure}
\centering\includegraphics[height=8cm,clip]{\FIGDIR/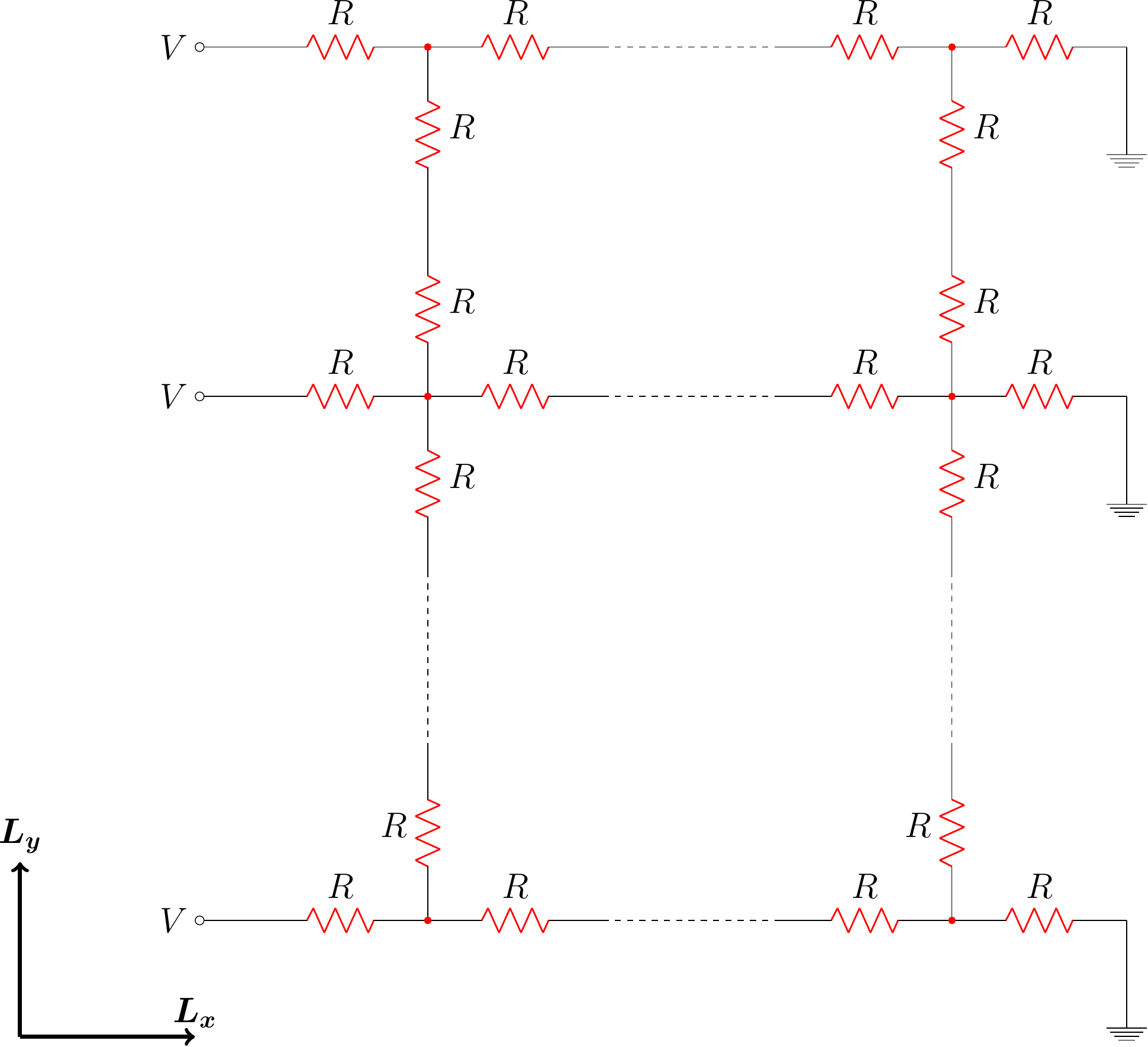}
\caption{A resistor network on a rectangular lattice geometry.}
 \label{fig:binary:res:network}
\end{figure}

As the most common 2D geometry, we begin with a finite size rectangular lattice defined by lengths $L_x$ and $L_y$.  Following the setup defined in the previous section,  a voltage $V$ is applied at one end of the lattice, in the direction of the length $L_x$ (see ~\fref{fig:binary:res:network}) while the other end is grounded. Resistances, each with value $R$, are connected to each site in all four directions. Our objective is to find out the effective resistance ($\Reff$) for the geometry and how $\Reff$ depends on the dimensions $L_x$ and $L_y$. 
%
%
%
%
\subsection{Analytical solution:}
\begin{figure}[!htp]
\centering
 \includegraphics[height=7cm,clip]{\FIGDIR/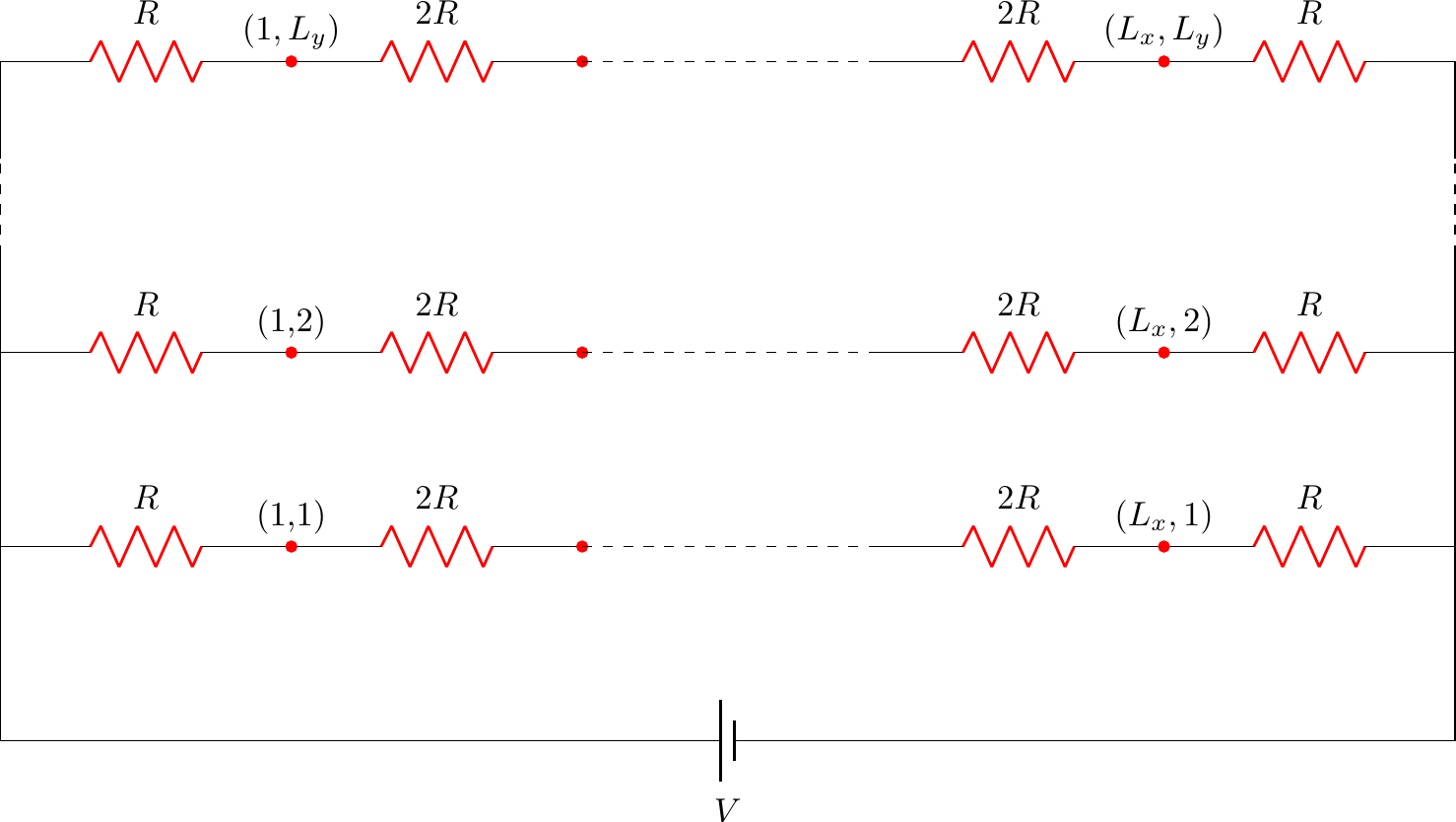}
 \caption{A network of resistors connected only in the $x$-direction.}
 \label{fig:R:only:in:xdir}
\end{figure}

To find out the effective resistance in the lattice, for a moment, we
assume there are no resistive connections in the $y$ (vertical) direction.    
Thus for a lattice of size $L_x \times L_y$, points are connected only in the $x$-direction (see ~\fref{fig:R:only:in:xdir}). There are total $L_y$ branches of parallel resistances with each branch consisting of a set of resistances in series. Now in each set of resistances in series, we can notice that the equivalent resistance between two adjacent sites is  $R+R=2R$ (resistance $R$ on the left of one site and on the right of the other site). Thus we find $L_x-1$ number of resistances of value $2R$ between first and last lattice points and two resistances of value $R$ on the left and right ends of the lattice. Thus in total, we have $L_x$ number of resistances of value $2R$ in series on each branch. Thus equivalent resistance of each branch is $2R L_x$. Since there exist $L_y$ such branches in parallel, the overall effective resistance of this simplified circuit:  
\begin{align}
R_{\eff}^\simp=\big[1/(2R L_x)+\cdots\mbox{($L_y$ times)}\big]\inv=2RL_x/L_y\,.
\label{eq:Reff:simp:rect}
\end{align}
%

\begin{figure}[!htp]
    \centering
      \subfigure[]{
     \includegraphics[height=3cm,clip]{\FIGDIR/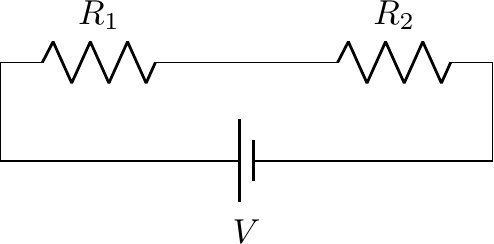}
     \label{fig:2res:in:series}
      }
      \subfigure[]{
     \includegraphics[height=4cm,clip]{\FIGDIR/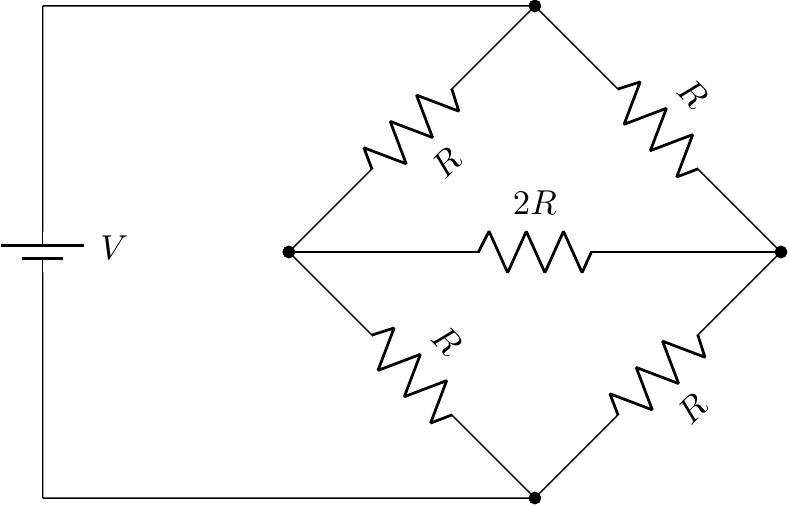}
     \label{fig:WB}
      }
       \caption{(a) Two resistances in series. (b) A simple Wheatstone bridge, which is a realization of the case $L_x=1$, $L_y=2$ in the rectangular network.}
\label{fig:rect:simplified}
\end{figure}
%
We know that when two resistances ($R_1$ and $R_2$) are connected in series, as shown in ~\fref{fig:2res:in:series}, the potential drop after the first resistance $R_1$ (i.e. in the middle of $R_1$ and $R_2$) will be given by 
\blgn
  V-V_1=V\frac{R_1}{R_1+R_2}\,, 
\elgn
which gives the potential at the middle of $R_1$ and $R_2$:
\blgn
V_1=V\bigg(1+\f{R_1}{R_1+R_2}\bigg)\,.
\elgn

Extending this argument to our case, the potential at a point $(i,j)$ will be  
\blgn
   V_{i,j}=V\,\bigg(1+\frac{2iR}{2RL_x}\bigg)=V\,\bigg(1+\f{i}{L_x}\bigg)\,.
\label{eq:pot:at:ij}
\elgn
since there are equivalent resistance of value $(i-1).2R+R=2iR$ [$(i-1)$ resistances with value $2R$ plus a single resistance with value $R$] resistances to the left of point $(i,j)$. \eref{eq:pot:at:ij} shows that the potential at any branch is independent of $y$-coordinate.

Thus, even if we were to connect the points in the $y$-direction using resistances of the same value (which was our original network to begin with), no current would flow in the $y$-direction for the same $x$-coordinate. This means that the original network, with all the lattice points joined, is equivalent to the network with lattice points joined only in the $x$-direction. Since the two networks are equivalent, the effective resistances of the original rectangular network will be the same as the one in \eref{eq:Reff:simp:rect}:
\blgn
R_\eff^\rect = 2R\,\frac{L_x}{L_y} = R\f{z}{2}\frac{L_x}{L_y}\,
\label{eq:Reff:rect:ana}
\elgn
where we attempt to write the formula in a more generic form by looking at the coordination number $z$ (number of nearest neighbor sites, $z=4$ for a rectangular lattice). We can easily notice that a balanced Wheatstone bridge~\cite{book:bird10} with resistance $R$ on each of its branches is the $L_x=1$ and $L_y=2$ case of the rectangular network (see \fref{fig:WB}). There, by applying \eref{eq:Reff:rect:ana}, we get $\Reff=2R.1/2=R$ which is supposed to be the desired result for the bridge network.

\subsection{Numerical Formulation:}
In our rectangular lattice of size $L_x\times L_y$, we can mark out distinct 9 kinds of lattice points:
\begin{itemize}
 \item Left bottom corner point ($i=1$, $j=1$)
 \item Left top corner point ($i=1$, $j=L_y$)
 \item Right bottom corner point ($i=L_x$, $j=1$)
 \item Right top corner point ($i=L_x$, $j=L_y$)
 \item Left Non-corner edge points ($i=1$, $j\in[2,L_y-1]$)
 \item Bottom non-corner edge points ($i\in[2,L_x-1]$, $j=1$)
 \item Right non-corner edge points ($i=L_x$, $j\in[2,L_y-1]$)
 \item Top non-corner edge points ($j=L_y$, $i\in[2,L_x-1]$)
 \item Non-border inner points ($i\in[2,L_x-1]$, $j\in[2,L_y-1]$)
\end{itemize}
The KCLs for the above 9 kinds of points follow:
\begin{enumerate}
%
\item \textbf{Left bottom corner point \ra} $i = 1$, $j = 1$:
\blgn
\frac{V − V_{1,1}}{R} + \frac{V_{2,1} − V_{1,1}}{2R} + \frac{V_{1,2} − V_{1,1}}{2R}= 0\,.
\elgn
\item \textbf{Left top corner point \ra} $i = 1$, $j = L_y$:
\blgn
\frac{V − V_{1,Ly}}{R}+\frac{V_{2,Ly} − V_{1,Ly}}{2R}+\frac{V_{1,Ly−1} − V_{1,Ly}}{2R}= 0\,.
\elgn
\item \textbf{Right bottom corner point \ra} $i = L_x$, $j = 1$:
\blgn\frac{V_{Lx−1,1} − V_{Lx,1}}{2R}-\frac{V_{Lx,1}}{R}+\frac{V_{Lx,2} − V_{Lx,1}}{2R}= 0\,.
\elgn
\item \textbf{Right top corner point \ra} $i = L_x$, $j = L_y$:
\blgn\frac{V_{Lx−1,Ly} − V_{Lx,Ly}}{2R}-\frac{V_{Lx,Ly}}{R}+\frac{V_{Lx,Ly−1} − V_{Lx,Ly}}{2R}= 0\,.
\elgn
%
\item \textbf {Left non-corner edge point \ra} $i = 1$, $j = 2$ to $L_y − 1$:
\blgn 
\frac{V − V_{1,j}}{R}+\frac{V_{2,j} − V_{1,j}}{2R}+\frac{V_{1,j−1} − V_{1,j}}{2R}+\frac{V_{1,j+1} − V_{1,j}}{2R}= 0\,.
\elgn
\item \textbf{Right non-corner edge point \ra} $i = L_x$, $j = 2$ to $L_y − 1$:
\blgn
\frac{V_{Lx−1,j} − V_{Lx,j}}{2R}-\frac{V_{Lx,j}}{R}+\frac{V_{Lx,j−1} − V_{Lx,j}}{2R}+\frac{V_{Lx,j+1} − V_{Lx,j}}{2R}= 0\,.
\elgn
\item \textbf{Bottom non-corner edge point \ra} $i = 2$ to $L_x$, $j = 1$:
\blgn
\frac{V_{i−1,1} − V_{i,1}}{2R}+\frac{V_{i+1,1} − V_{i,1}}{2R}+\frac{V_{i,2} − V_{i,1}}{2R}= 0\,.
\elgn
\item \textbf{Top non-corner edge point \ra} $i = 2$ to $L_x$, $j = L_y$:
\blgn
\frac{V_{i−1,Ly} − V_{i,Ly}}{2R}+\frac{V_{i+1,1} − V_{i,Ly}}{2R}+\frac{V_{i,Ly−1} − V_{i,Ly}}{2R}= 0\,.
\elgn
%
\item \textbf{Non-border inner point \ra} $i = 2$ to $L_x − 1$, $j = 2$ to $L_y − 1$:
\blgn
\frac{V_{i−1,j} − V_{i,j}}{2R}+\frac{V_{i+1,j} − V_{i,j}}{2R}+\frac{V_{i,j−1} − V_{i,j}}{2R}+\frac{V_{i,j+1} − V_{i,j}}{2R}= 0\,.
\elgn
\end{enumerate}

We can rearrange the above equations by collecting 
the coefficients of $V_{ij}$:
\begin{enumerate}
\item  {\bf Left bottom corner point \ra} $i=1$, $j=1$:
\be
\bl\f{1}{R}+\f{1}{2R}+\f{1}{2R}\br
V_{1,1}
-\f{1} {2R}V_{2,1}
-\f{1}{2R}V_{1,2}
=\f{V}{R}\,. 
\label{eq:KCL:rect:rearranged:1st}
\ee
\item   {\bf Left top corner point \ra} $i=1$, $j=L_y$:
\blgn
\bl
\f{1} {R}+\f{1}{2R }
+\f{1}{2R}
\br V_{1,L_y}
-\f{1} {2R} V_{2,L_y}
-\f{1} {2R} V_{1,L_y-1}
=\f{V} {R}\,.
\label{eq:KCL:rect:rearranged:2nd}
\elgn
\item  {\bf Right  bottom corner point \ra} $i=L_x$, $j=1$:
\blgn
\bl
&\f{1}{2R}+\f{1}{R}
+\f{1}{2R}
\br V_{L_x,1}-\f{1 }{2R}V_{L_x-1,1}
-\f{1} {2R}V_{L_x,2}
=0\,.
\elgn
\item  {\bf Right top corner point \ra} $i=L_x$, $j=L_y$:
\blgn
\bl
\f{1}{2R}
+\f{1}{R}
+\f{1} {2R}
\br V_{L_x,L_y} 
-\f{1}{2R} V_{L_x-1,L_y}
-\f{1}{2R} V_{L_x,L_y-1}
=0\,.
\elgn
\item{\bf Left non-corner edge point \ra} $i=1$, $j=2$ to $L_y-1$:
\blgn
\bl
\f{1}{R} +\f{1}{2R}
+\f{1}{2R}+\f{1}{2R}
\br V_{1,j}
-\f{1} {2R}V_{2,j}
-\f{1 } {2R}V_{1,j-1}
-\f{1 } {2R}V_{1,j+1}
=\f{V}{R}\,.
\elgn
\item  {\bf Right non-corner edge point \ra} $i=L_x$, $j=2$ to $L_y-1$:
\blgn
\bl
\f{1}{2R}
+\f{1}{R}
+\f{1} {2R}
+\f{1} {2R}
\br V_{L_x,j}
-\f{1 } {2R}V_{L_x-1,j}
-\f{1} {2R}V_{L_x,j-1}
-\f{1 } {2R}V_{L_x,j+1}
=0\,.
\elgn
\item  {\bf Bottom non-corner edge point \ra} $i=2$ to $L_x$, $j=1$:
\blgn
\bl 
\f{1} {2R }
+\f{1}{2R}
+\f{1}{2R }
\br V_{i,1}
-\f{1}{2R} V_{i-1,1}
-\f{1}{2R} V_{i+1,1}
-\f{1 }{2R} V_{i,2}
=0\,.
\elgn
\item  {\bf Top non-corner edge point \ra} $i=2$ to $L_x$, $j=L_y$:
\blgn
\bl
\f{1}{2R}
+\f{1}{2R}
+\f{1}{2R}
\br V_{i,L_y}
-\f{1 } {2R}V_{i-1,L_y}
-\f{1} {2R}V_{i+1,1}
-\f{1}  {2R}V_{i,L_y-1}
=0\,.
\elgn
%
%
\item  {\bf Non-border inner point \ra} $i=2$ to $L_x-1$, $j=2$ to $L_y-1$:
\blgn
\bl
&\f{1}{2R}
+\f{1}{2R}
+\f{1} {2R}
+\f{1}{2R}
\br V_{i,j}
-\f{1 } {2R}V_{i-1,j}
-\f{1 } {2R}V_{i+1,j}
-\f{1 } {2R}V_{i,j-1}
-\f{1} {2R}V_{i,j+1}
=0\,.
\label{eq:KCL:rect:rearranged:last}
\elgn
\end{enumerate}
Now $V_{ij}$'s constitute a $L_x \times L_y$ matrix, but if we linearize (see Appendix for details) it as a column
vector $\bf V$ of length $L_x L_y$, 
The above equations can be represented in matrix notation as 
\blgn 
{\bf GV=I}
\elgn
where $\bf I$ is a column vector whose values are given by the right hand side of equations  $(7)$ to $(15)$ and \textbf{G} is a matrix consisting of the coefficients of the variables in the equations. Since \textbf{G} has units $1/R$, we are calling it the \textit{conductance matrix}. Since the potential at any lattice point $(i,j)$ depends only on its neighboring points, the matrix \textbf{G} is generally sparse, and can be solved using a sparse matrix solver numerically. The method is very similar to typical transfer matrix method used in circuit analysis~\cite{book:omalley:schaumseries11}  and also similar to the method used in the context of disordered resistor network~\cite{derrida:vannimenus:jpamag82}.

Once the above matrix system is solved and we know the potential at all lattice points, the effective resistance can be determined by dividing the total applied voltage by net current flowing through the lattice in the direction of the applied voltage. It can be observed that the net current can be determined using the potential of the end-points of the lattice. The net current, in this case, would be given by 
\blgn 
I_{\net}=\frac{\Sigma^{L_y}_{j=1}V_{Lx,j}}{R}\,. 
\elgn

Effective Resistance can then be determined as 
\blgn 
R_{\eff}=\frac{V}{I_{\net}}\,.
\elgn
In practice, we choose $R=1$ and $V=1$.

%
%
\subsection{Results:}
\begin{figure}[H]
 \centering
 \subfigure[]{
 \includegraphics[height=5.5cm,clip]{\FIGDIR/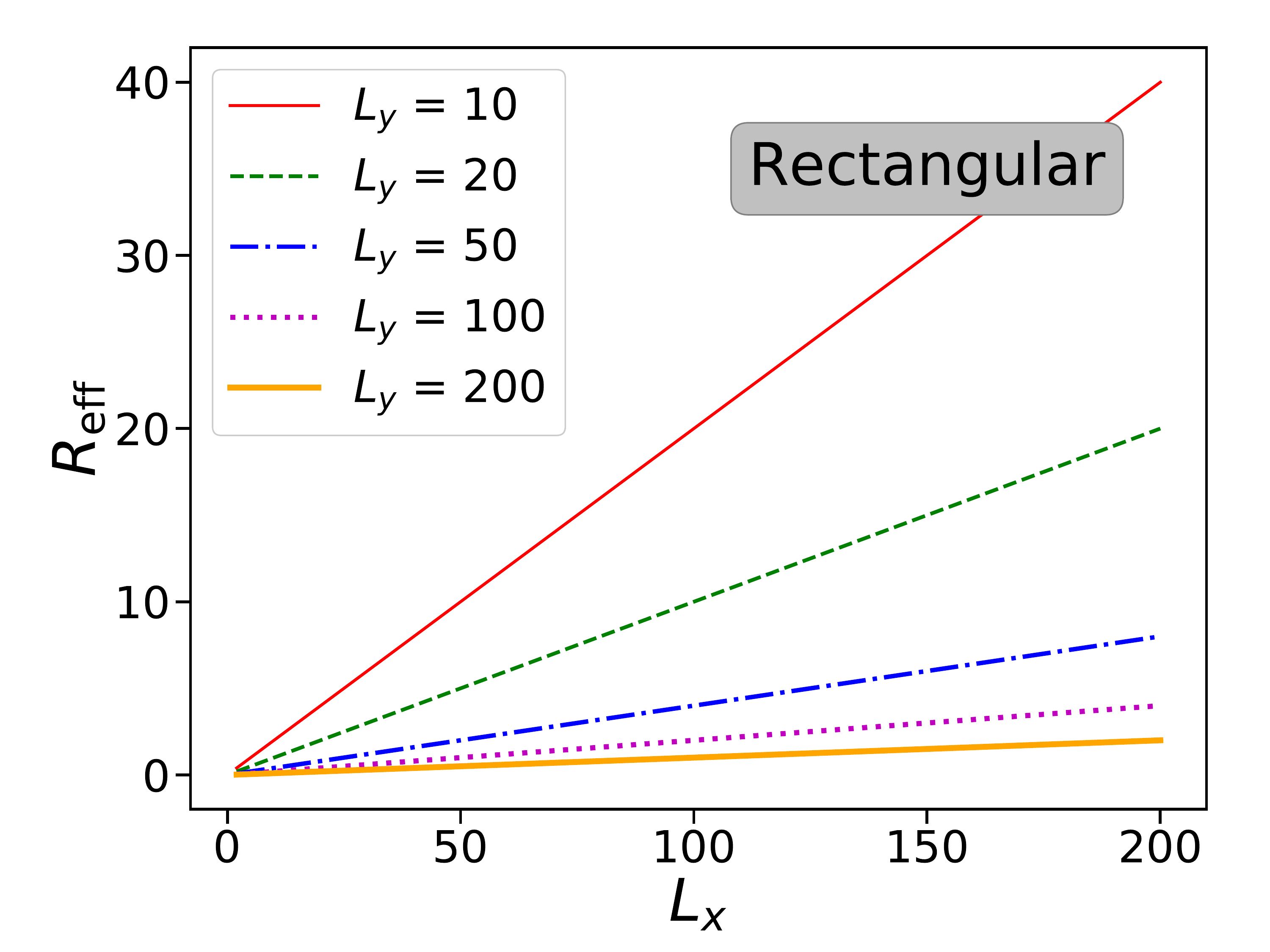}
 \label{fig:Reff:vs:Lx:rect}
 }
 \subfigure[]{
 \includegraphics[height=5.5cm,clip]{\FIGDIR/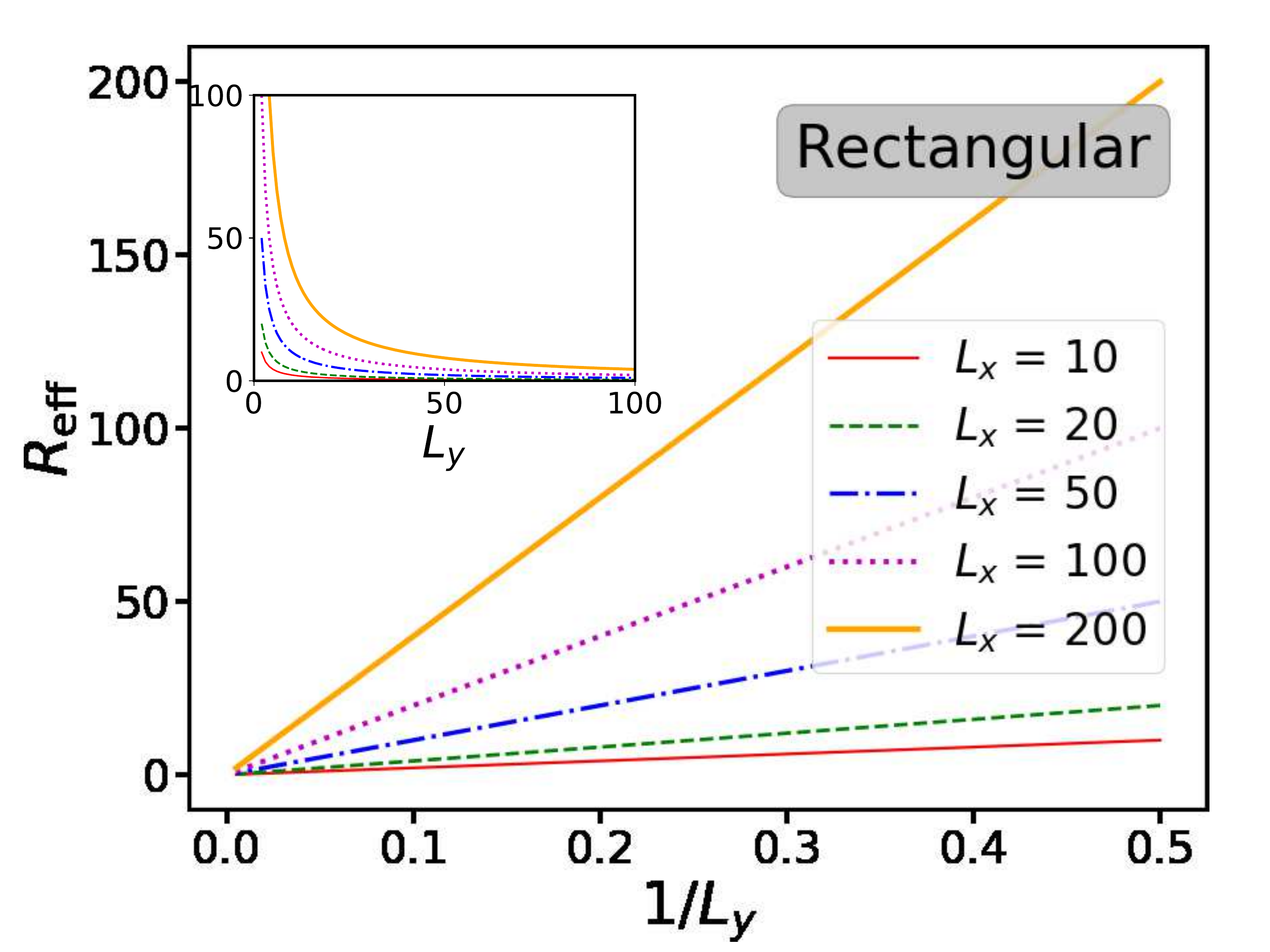}
 \label{fig:Reff:vs:Ly:rect}
 }
 \caption{A plot of $R_{\eff}$ as (a) $L_x$ is varied for different values of $L_y$ and (b) $L_y$ is varied for different values of $L_x$ for a rectangular lattice.}
 \end{figure}

\begin{figure}[!hbp]
\centering
 \subfigure[]{
\includegraphics[height=5.5cm]{\FIGDIR/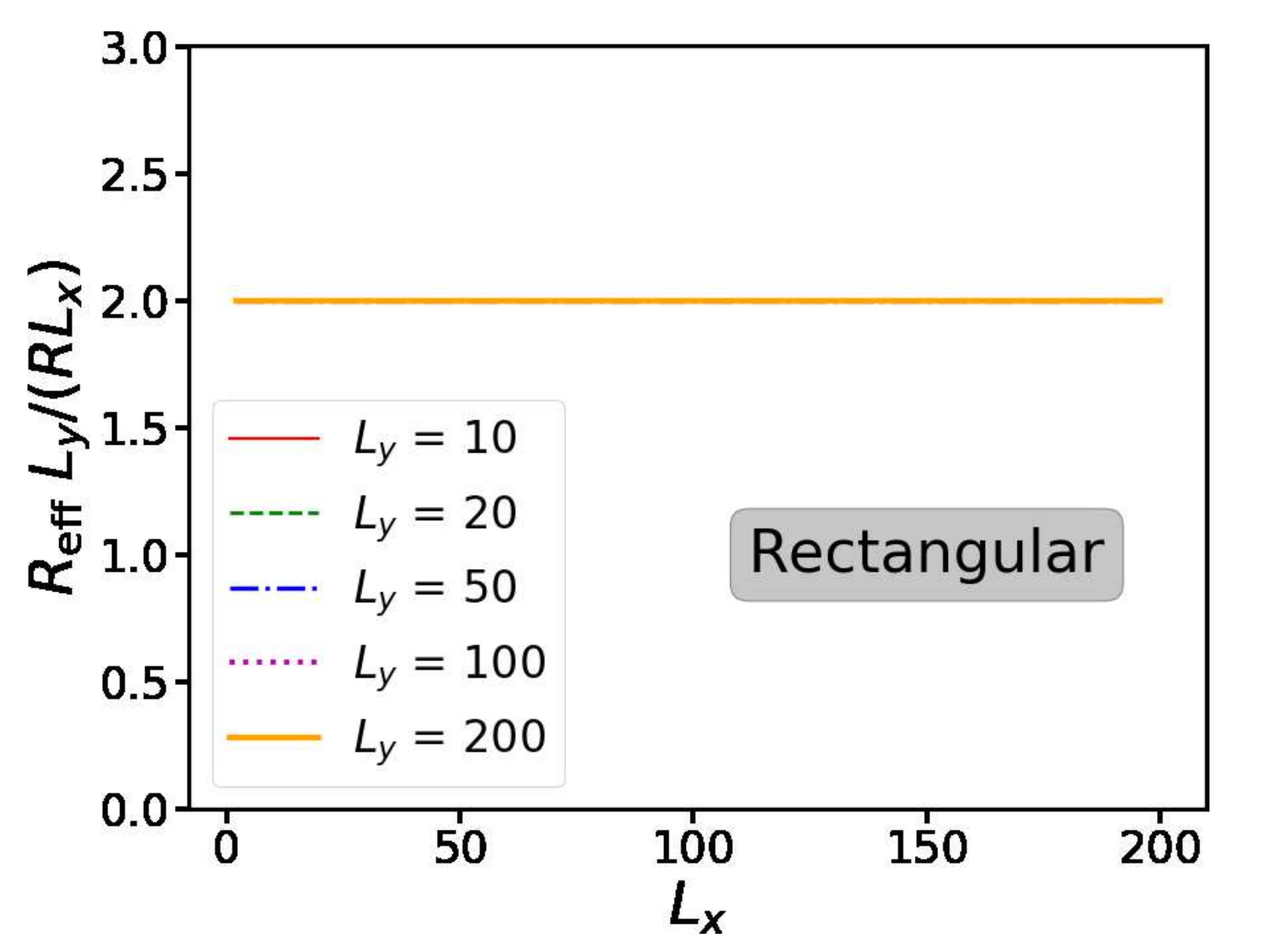}
\label{fig:Rratio:vs:Lx:rect}
}
 \subfigure[]{
\includegraphics[height=5.5cm]{\FIGDIR/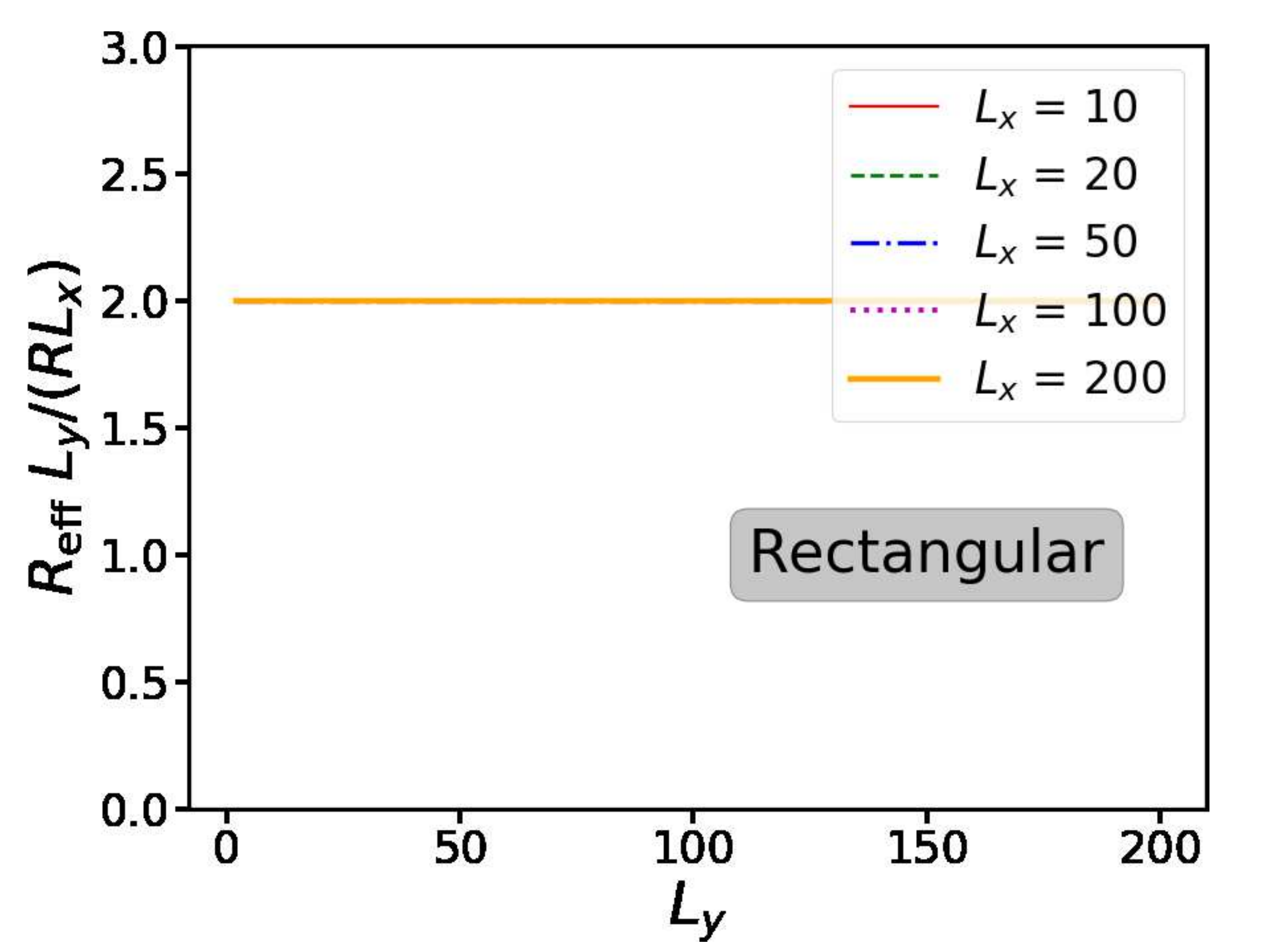}
\label{fig:Rratio:vs:Ly:rect}
}
\caption{A plot of $r=\Reff L_y/ (R L_x)$ as (a) $L_x$ and (b) $L_y$ is varied while other parameters are kept fixed. Both show $R_\ratio=2$ and it is independent of dimensions $L_x$ and $L_y$.} 
\label{fig:Rratio:rect}
\end{figure}

We first plot $\Reff$ against $L_x$ for several fixed values of $L_y$. As expected from the
analytical solution expressed in \eref{eq:Reff:rect:ana}, $\Reff$ grows linearly as $L_x$ increases and the slope of the linear curve drops at a larger value of $L_y$ (see \fref{fig:Reff:vs:Lx:rect}).  
%
When $L_x$ is kept constant, $\Reff$ decreases as $L_y$ increases and $\Reff$ vs $1/L_y$ plots show linear, establishing that $\Reff \propto L_x/L_y$. Now to find out the proportionality constant,
we define 
\blgn
r \equiv \f{\Reff L_y}{R L_x}
\elgn
which according to \eref{eq:Reff:rect:ana} should be equal to $z/2=2$. Both \fref{fig:Rratio:vs:Lx:rect} and \fref{fig:Rratio:vs:Ly:rect} show that $r$ is a constant when $L_x$ and $L_y$ are varied respectively, keeping the other dimension as a fixed parameter. The value of the constant is 2 and hence we see that the numerical results very well agree with our analytical formula.

\section{Hexagonal Network Model}
\label{sec:hex:network}
Now we consider the hexagonal or the graphene~\cite{neto:etal:rmp09} type honeycomb lattice network. Out of two possible orientations, we select a hexagonal lattice which has armchair edges in the $x$-direction and zigzag edges in the $y$-direction
(see \fref{fig:armhex:rn}) and dub this armchair hexagonal lattice.
  
\subsection{Numerical formulation:}  
Here for our convenience, we break the sites into two categories -- (i) $M$-type sites, sitting at the middle corners of a hexagon and such sites connect to the bias and grounding,  and (ii) $S$-type sites, sitting on the top or bottom sides of a hexagon. We add extra indices 0 and 1 to specify $M$ and $S$ sites respectively. Now we can see there must be always equal and even numbers of $M$ and $S$ sites in the $x$-direction in a lattice with complete hexagons. The number of  $M$ and $S$ sites ($L_M$ or $L_S$)  
sets the measurement of the length $L_x$: $L_x=L_x^M=L_x^S$. On the other hand, the number of voltage connection determines the length $L_y$: $L_y=L_y^M$, $L_y^S=L_y^M+1=L_y+1$. Total number of sites can be determined as $N_\site=L_x^M L_y^M + L_x^S L_y^S = L_x L_y + L_x (L_y+1)=L_x(2L_y+1)$. 
We can distinguish 6 kinds of sites in this system: 
%
%
%
\begin{figure}
\centering\includegraphics[height=9cm,clip]{\FIGDIR/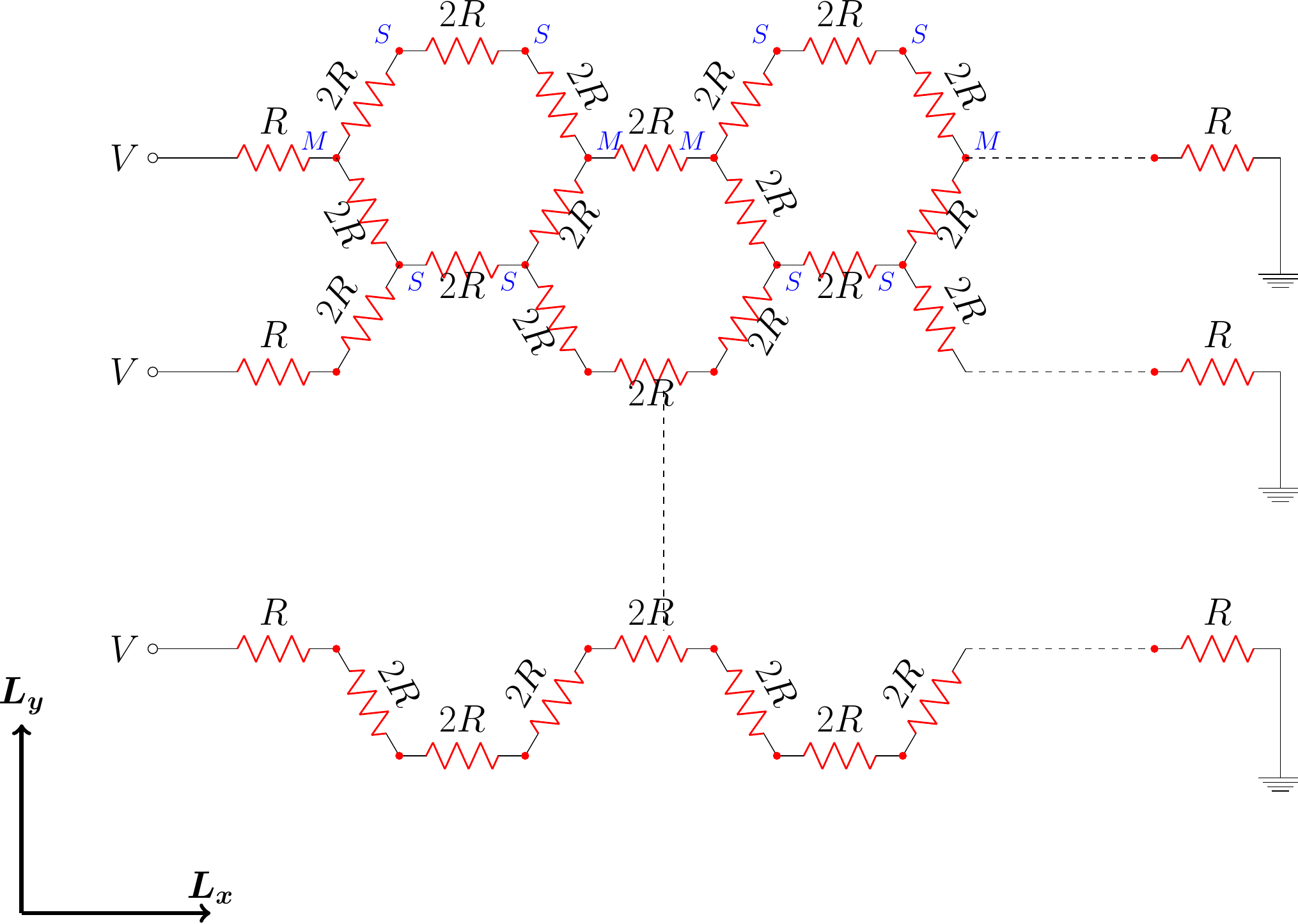}
\caption{A resistor network on a hexagonal lattice geometry.}
 \label{fig:armhex:rn}
\end{figure}

\begin{itemize}
\item Left Border Points ($i=1$, $j=1$ to $L_y$, $k=0$)
\item Right Border Points ($i=L_x$, $j=1$ to $L_y$, $k=0$) 
\item Top border points ($i=1$ to $L_x,j=L_y+1,k=1$)
\item Bottom Border Points  ($i=1$ to $L_x$, $j=1$, $k=1$)
\item $M$-type inner points ($i=2$ to $L_x-1$, $j=1$ to $L_y$, $k=0$) 
\item $S$-type inner points ($i=1$ to $L_x$, $j=2$ to $L_y$, $k=1$) 
\end{itemize}

The KCL for the above 6 kinds of points would be:
\begin{enumerate}
\item \textbf{Left Border Points \ra} $i=1$, $j=1$ to $L_y$, $k=0$:
\blgn 
\frac{V-V_{1,j,0}}{R}+\frac{V_{1,j,1}-V_{1,j,0}}{2R}+\frac{V_{1,j+1,1}-V_{1,j,0}}{2R}=0\,. 
\elgn
\item \textbf{Right Border Points \ra} $i=L_x$, $j=1$ to $L_y$, $k=0$:
\blgn 
\frac{-V_{2Lx,j,0}}{R}+\frac{V_{2Lx,j,1}-V_{2Lx,j,0}}{2R}+\frac{V_{2Lx,j+1,1}-V_{2Lx,j,0}}{2R}=0\,.
\elgn
\item \textbf{Top Border Points \ra} $i=1$ to $L_x$, $j=L_y+1$, $k=1$:\\
if i=odd, 
\blgn 
\frac{V_{i,Ly,0}-V_{i,Ly+1,1}}{2R}+\frac{V_{i+1,Ly+1,1}-V_{i,Ly+1,1}}{2R}=0\,. 
\elgn
if i=even, 
\blgn 
\frac{V_{i,Ly,0}-V_{i,Ly+1,1}}{2R}+\frac{V_{i-1,Ly+1,1}-V_{i,Ly+1,1}}{2R}=0\,. 
\elgn
\item \textbf{Bottom Border Points \ra} $i=1$ to $L_x$, $j=1$, $k=1$:\\ 
if i=odd, 
\blgn 
\frac{V_{i,1,0}-V_{i,1,1}}{2R}+\frac{V_{i+1,1,1}-V_{i,1,1}}{2R}=0\,. 
\elgn
if i=even, 
\blgn 
\frac{V_{i,1,0}-V_{i,1,1}}{2R}+\frac{V_{i-1,1,1}-V_{i,1,1}}{2R}=0\,. 
\elgn
\item \textbf{$M$-type inner points \ra} $i=2$ to $L_x-1$, $j=1$ to $L_y$, $k=0$:\\ 
if i=odd, 
\blgn 
\frac{V_{i-1,j,0}-V_{i,j,0}}{2R}+\frac{V_{i,j,1}-V_{i,j,0}}{2R}+\frac{V_{i,j+1,1}-V_{i,j,0}}{2R}=0\,. 
\elgn
if i=even, 
\blgn 
\frac{V_{i+1,j,0}-V_{i,j,0}}{2R}+\frac{V_{i,j,1}-V_{i,j,0}}{2R}+\frac{V_{i,j+1,1}-V_{i,j,0}}{2R}=0\,. 
\elgn
\item \textbf{$S$-type inner points \ra} $i=1$ to $L_x$, $j=2$ to $L_y$, $k=1$:\\ 
if i=odd, 
\blgn 
\frac{V_{i+1,j,1}-V_{i,j,1}}{2R}+\frac{V_{i,j,0}-V_{i,j,1}}{2R}+\frac{V_{i,j-1,0}-V_{i,j,1}}{2R}=0\,. 
\elgn
if i=even, 
\blgn \frac{V_{i-1,j,1}-V_{i,j,1}}{2R}+\frac{V_{i,j,0}-V_{i,j,1}}{2R}+\frac{V_{i,j-1,0}-V_{i,j,1}}{2R}=0\,. 
\elgn
\end{enumerate}
Note that here we have two types of sites, namely $M$ and $S$, and that though we use $(i,j)$ to denote the two-dimensional location different types of sites, a particular kind of site belongs to a particular type and hence one type's $i$ or $j$ should not coincide with another type's $i$ or $j$ and this distinction is taken care by the index $k$. As before, the above equations can be represented in matrix notation and are solved using a sparse matrix solver. 

\subsection{Results:}
\begin{figure}[!htp]
\centering
\subfigure[]{
\includegraphics[height=5.5cm,clip]{\FIGDIR/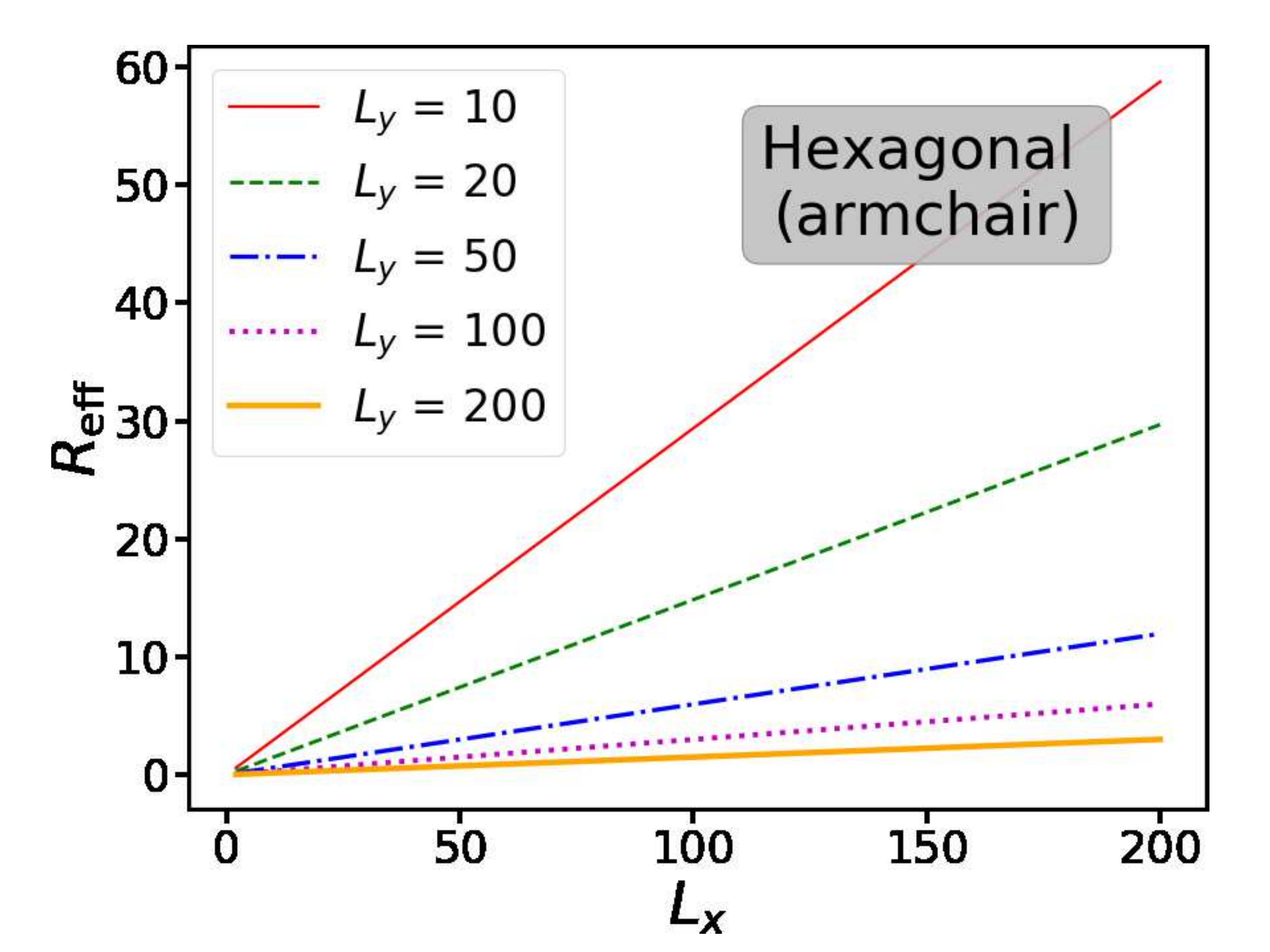}
\label{fig:Reff:vs:Lx:armhex}
}
\subfigure[]{
\label{fig:Reff:vs:Ly:armhex}
\includegraphics[height=5.5cm,clip]{\FIGDIR/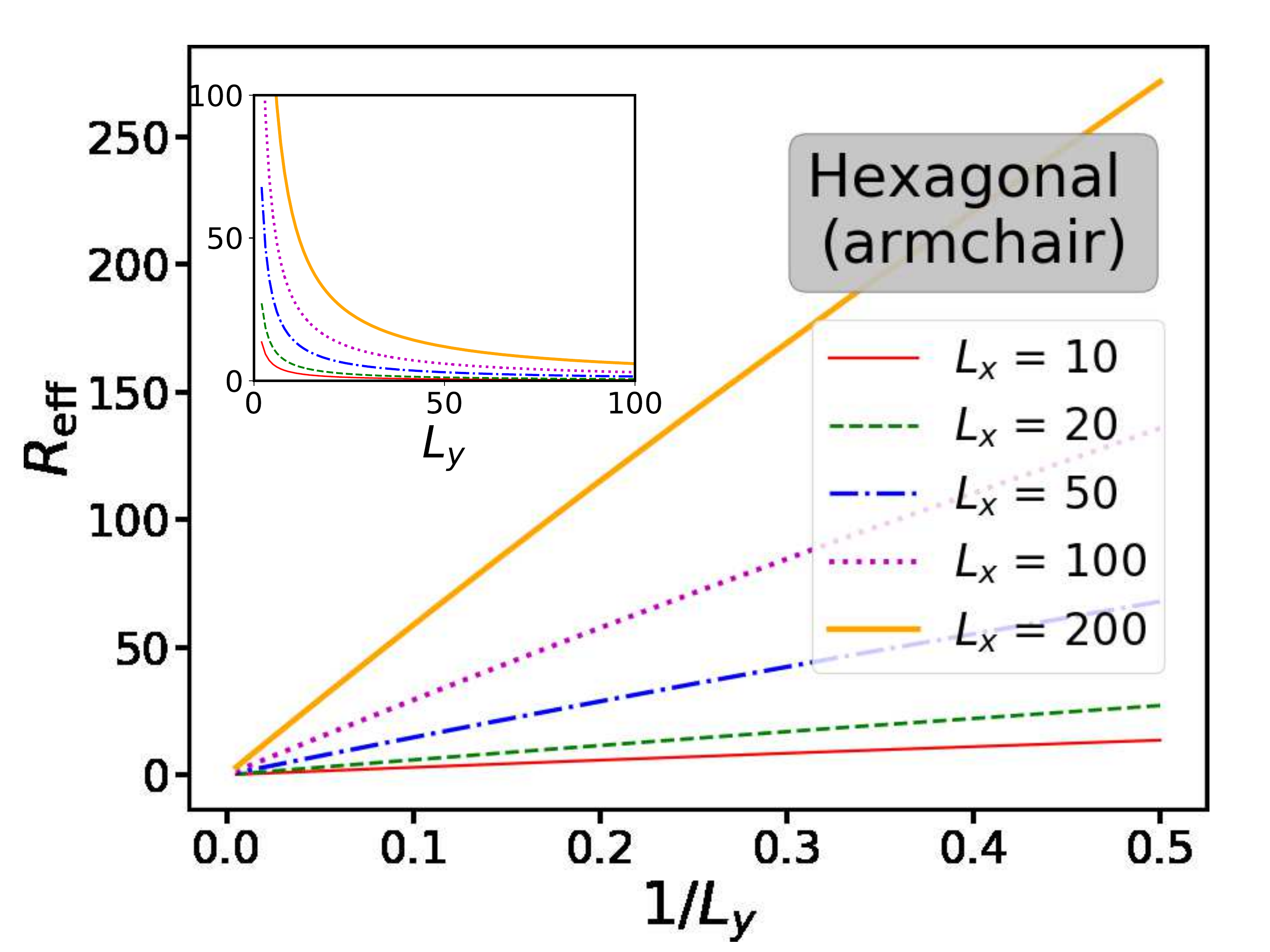}
}
\label{fig:Reff:armhex}
\caption{Plot of $R_{\eff}$ as (a)  $L_x$ is varied for different values of $L_y$ and (b) $L_y$ is varied for different values of $L_x$ for an armchair hexagonal lattice based network.}
\end{figure}

Like in the previous case, we first plot $\Reff$ as $L_x$ is varied, keeping $L_y$ fixed at different values. Even for a hexagonal lattice, $R_{\eff}$ seems to increase linearly with $L_x$, as seen in \fref{fig:Reff:vs:Lx:armhex}. We then plot $R_{\eff}$ as $L_y$ is varied, keeping $L_x$ fixed at different values. The result is shown in \fref{fig:Reff:vs:Ly:armhex}. Though $\Reff$ decreases with increasing $L_y$ like in the rectangular lattice case, $\Reff$ vs $1/L_y$ plots are not exactly linear (see ~\fref{fig:Reff:armhex}).

\begin{figure}[H]
\centering
\subfigure[]{
\includegraphics[height=6cm,clip]{\FIGDIR/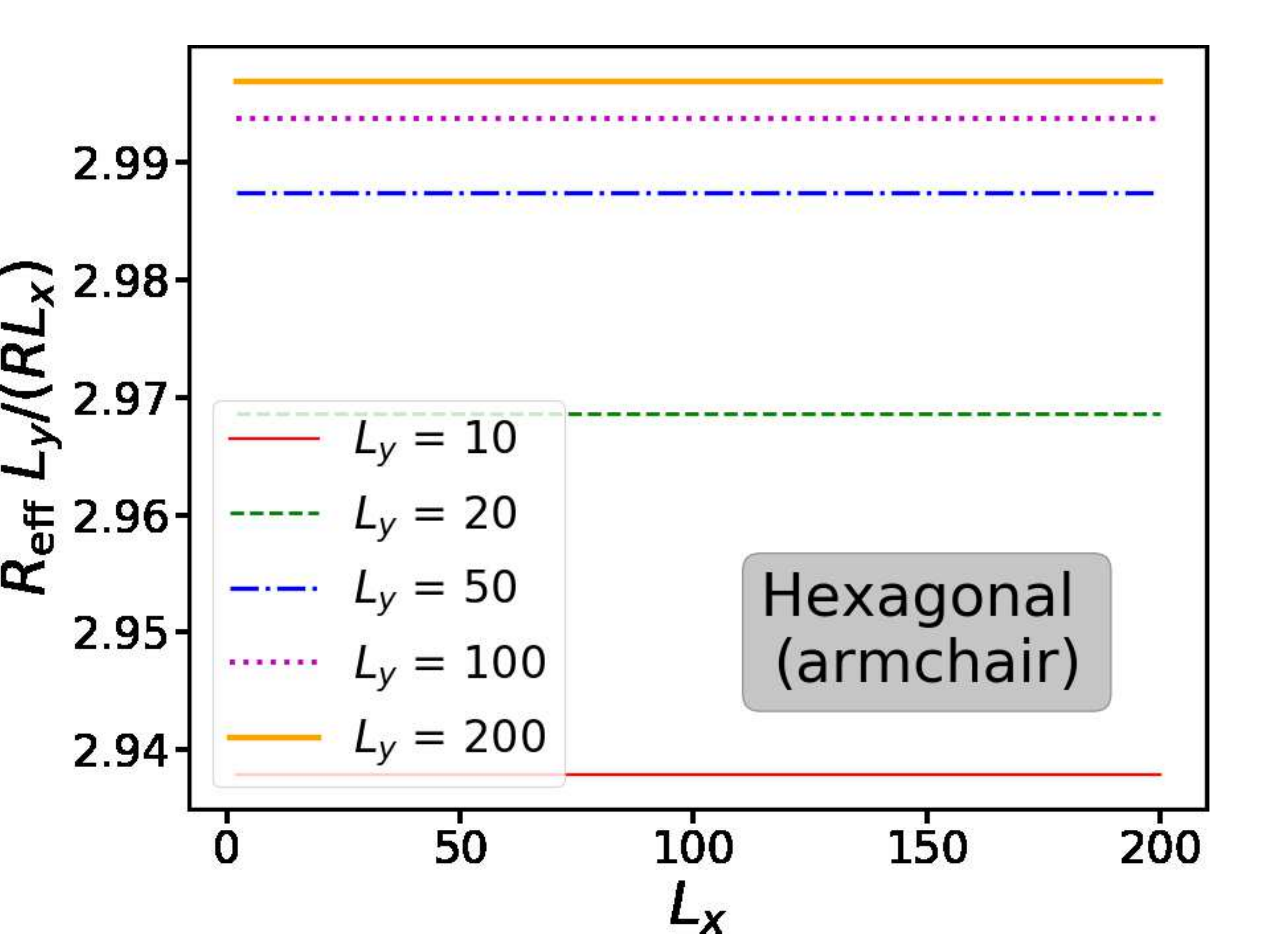}
\label{fig:Rratio:vs:Lx:armhex}
}
\subfigure[]{
\includegraphics[height=6cm,clip]{\FIGDIR/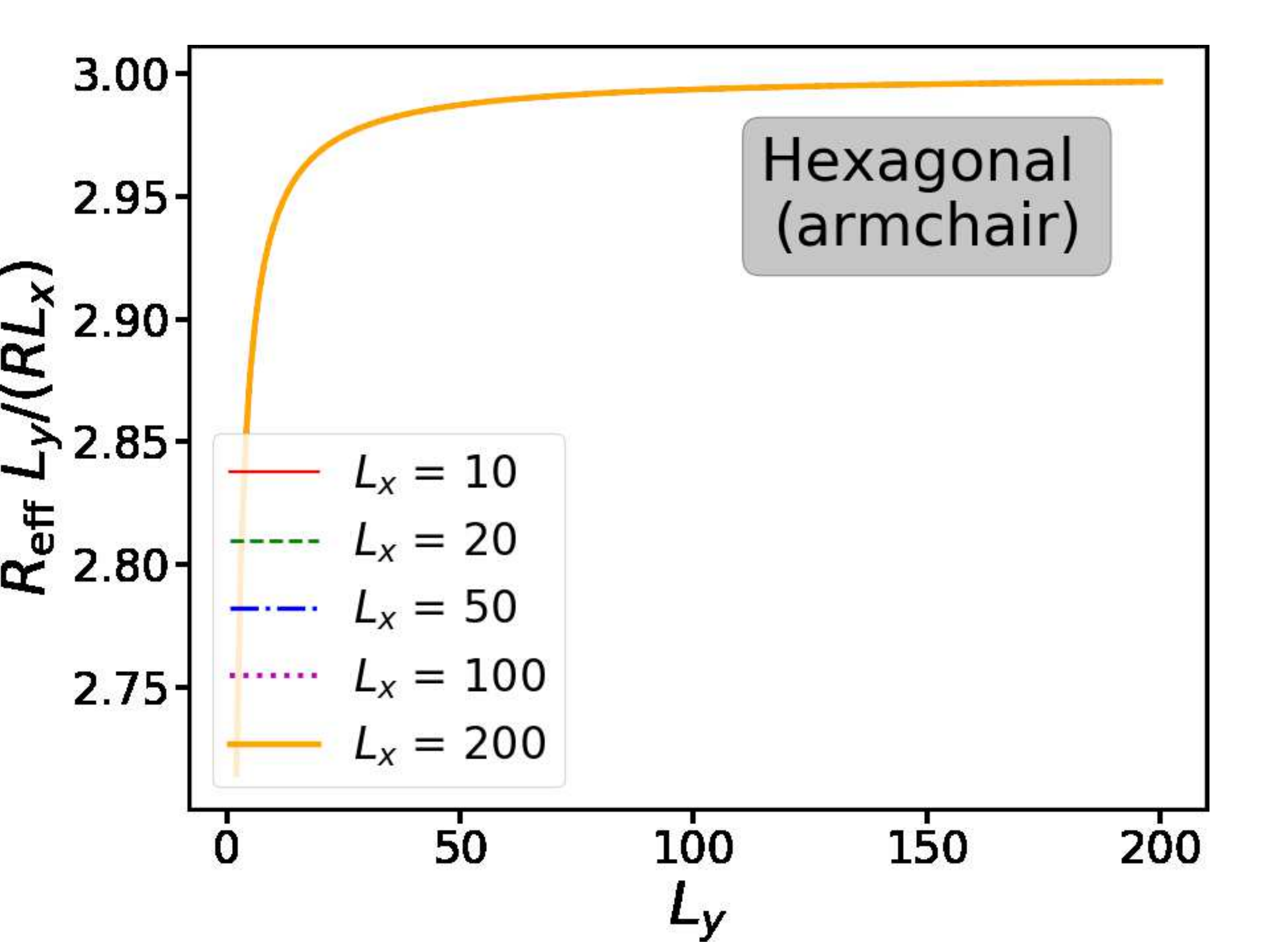}
\label{fig:Rratio:vs:Ly:armhex}
}
\subfigure[]{
\includegraphics[height=6cm,clip]{\FIGDIR/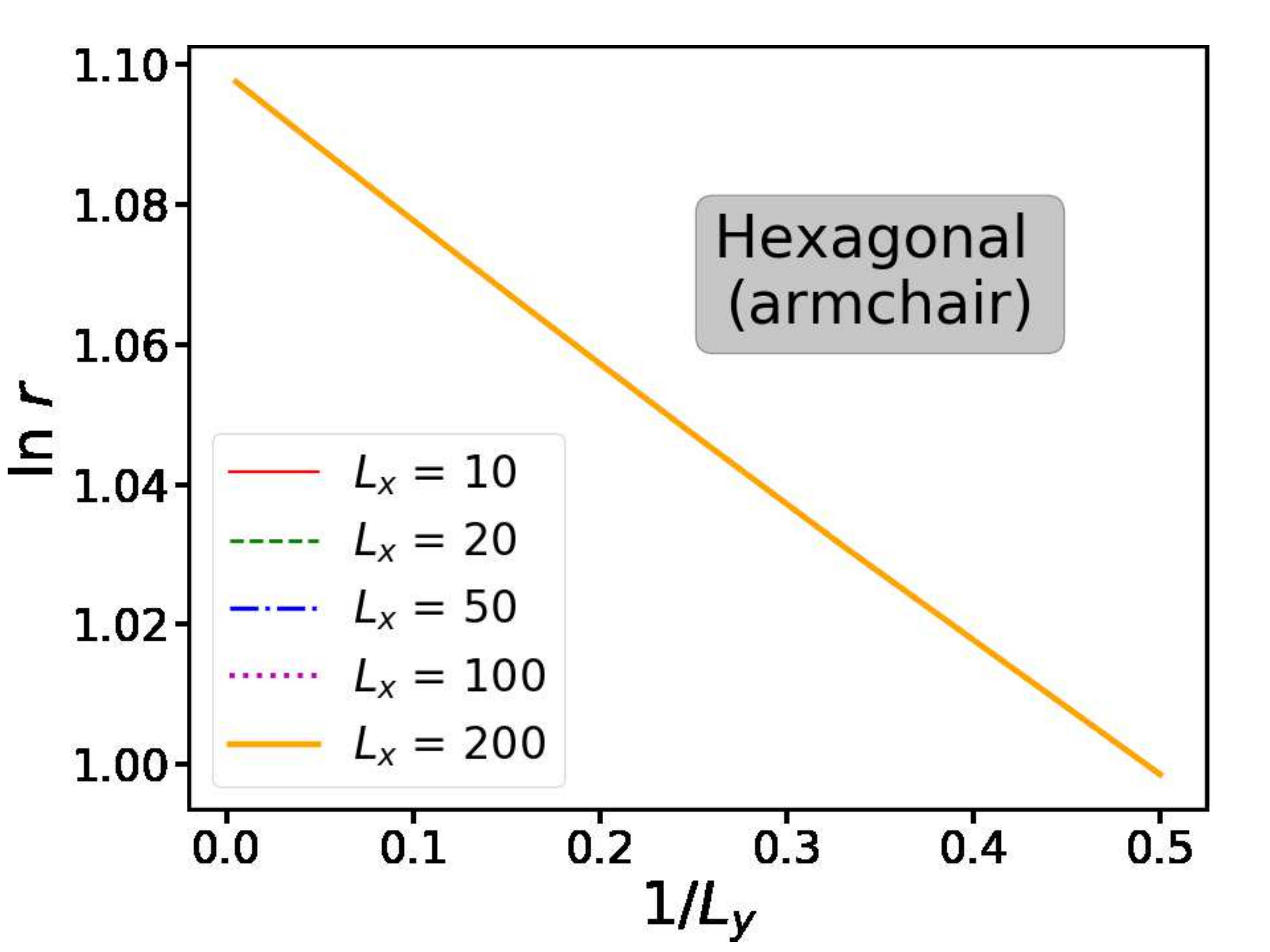}
\label{fig:lnRratio:vs:Lyinv:armhex}
}
\label{fig:Rratio:armhex}
\caption{Plot of $r=\Reff L_y/(R\L_x)$ as (a) $L_y$ is varied for different values of $L_x$ 
and (b) as $L_x$ is varied for different values of $L_y$
for an armchair hexagonal lattice. (c) $\ln r$ plotted against $1/L_y$ to verify the formula given in \eref{eq:al:guess} for $\al$ where $r=\al(z,L_y)$.}
\end{figure}

To see what is the actual dependence, we again plot the ratio $r=\Reff L_y/(RL_x)$ against $L_x$ and $L_y$ keeping other parameters fixed. We found a few interesting observations: 
(i) When $L_y$ is fixed, the ratio $r$ becomes independent of $L_x$; 
(ii) When $L_x$ is fixed, $r$ becomes universal and it approaches a constant at the thermodynamic limit ($L_y\to \infty$). These two observations, as shown in \fref{fig:Rratio:armhex}, let us arrive at a
conclusion that the ratio is a sole function of $L_y$:
\blgn
r=\alpha(L_y)\,.
\label{eq:r:armhex}
\elgn
This leads to an empirical formula for the effective resistance:
\blgn
\Reff^\hex=\al(L_y)\f{L_x}{L_y}\,.
\elgn
%
Now we further notice that $\al(z,L_y)$ approaches $z$ as $L_y\to \infty$ where $z=3$ is the lattice
coordination number for a hexagonal lattice. Since $\al(z,L_y)$ has to be dimensionless to keep \eref{eq:r:armhex} physically consistent, a convenient guess could be
\blgn
\al(L_y)=z\,e^{-c/L_y}\,
\label{eq:al:guess}
\elgn
which implies
\blgn
\ln \al(L_y)=\ln z-c L_y\inv\,. 
\elgn 
where $c$ is a constant. Now \fref{fig:lnRratio:vs:Lyinv:armhex} plots $\ln\al$ against $L_y\inv$
for various $L_x$ and we can see when $L_y\inv$ approaches zero (thermodynamic limit),  $\ln\al$ approaches $\ln z=\ln 3 \simeq 1.1$ vindicating our guessed formula for $\al$ in \eref{eq:al:guess}. 

%
%
%
%
%
%
\section{Triangular Network Model}
\begin{figure}
\centering\includegraphics[height=6cm,clip]{\FIGDIR/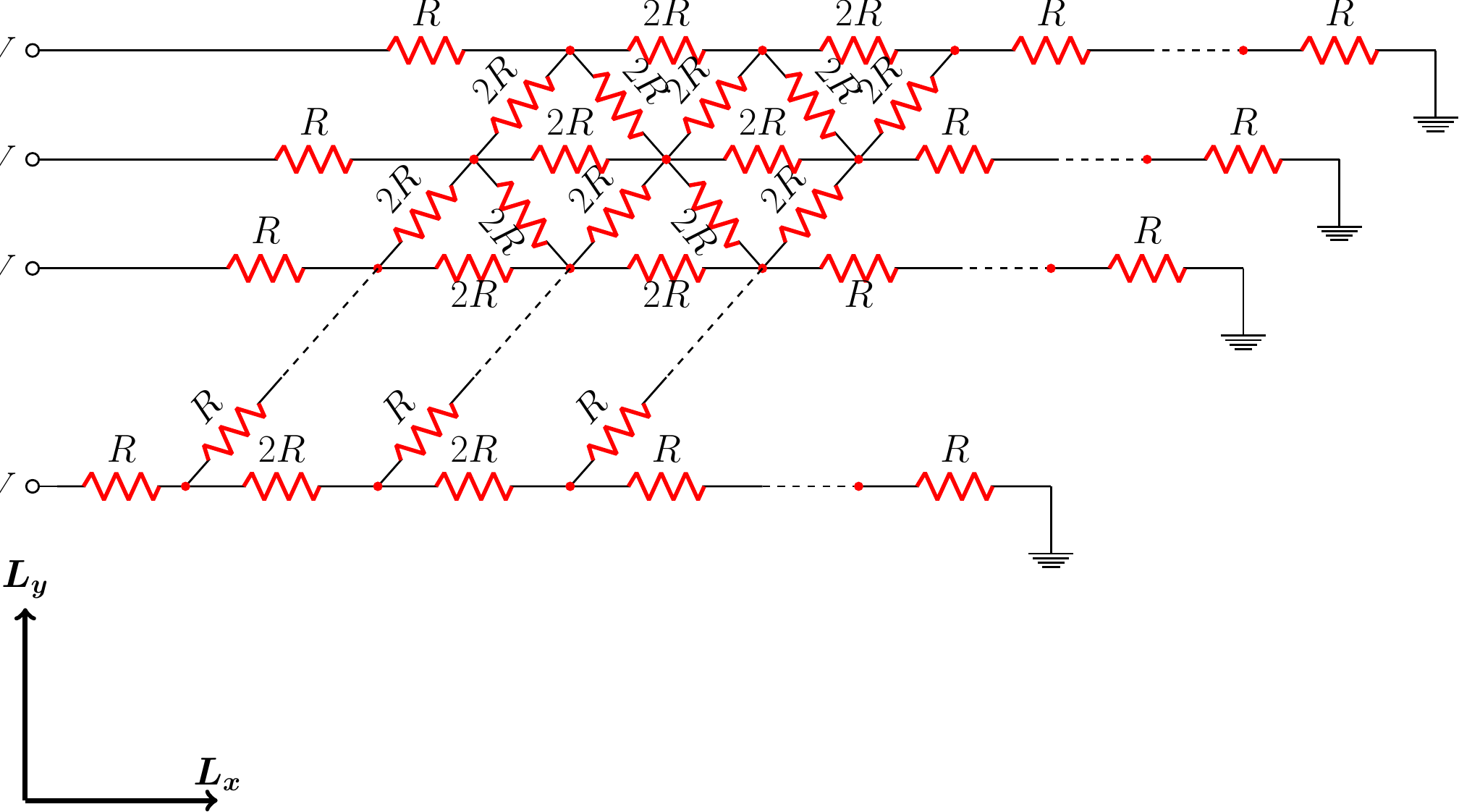}
\caption{A resistor network on a triangular lattice geometry.}
 \label{fig:tri:rn}
\end{figure}

We now move to the case where the lattice is triangular. The network considered is shown in \fref{fig:tri:rn}. Clearly this lattice is similar to a rectangular lattice, with the exception that diagonal points in one particular direction are also connected via an equivalent resistance $2R$.
\subsection{Numerical formulation:} 
For a triangular lattice, we have 9 kinds of lattice points:
\begin{itemize}
 \item Left bottom corner point $(i=1,j=1)$
 \item Left top corner point $(i=1,j=L_y)$
 \item Right bottom corner point $(j=1,i=L_x)$
 \item Right top corner point $(i=L_x,j=L_y)$
 \item Left non-corner edge points $(i=1, j\in[2,L_y-1])$
 \item Right non-corner edge points $(i=L_x,j\in[2,L_y-1])$
 \item Bottom non-corner edge points $(j=1, in\in[2,L_x-1])$
 \item Top non-corner edge points $(j=L_y,i\in[2,L_x-1])$
 \item Non-border inner points $(i\in[2,L_x-1],j\in[2,L_y-1])$)
\end{itemize}
The KCL, that relates the potential at any point $(i,j)$ with its neighboring point, for the above 9 kinds of points would be as follows:
\begin{enumerate}
\item \textbf{Left bottom corner point \ra} $i = 1$, $j = 1$:
\blgn
\frac{V − V_{1,1}}{R} + \frac{V_{2,1} − V_{1,1}}{2R} + \frac{V_{1,2} − V_{1,1}}{2R}= 0\,.
\elgn
\item \textbf{Left top corner point \ra} $i = 1$, $j = L_y$:
\blgn\frac{V − V_{1,Ly}}{R}+\frac{V_{2,Ly} − V_{1,Ly}}{2R}+\frac{V_{1,Ly−1} − V_{1,Ly}}{2R}+\frac{V_{2,Ly−1} − V_{1,Ly}}{2R}= 0\,.
\elgn
\item \textbf{Right bottom corner point \ra} $i = L_x$, $j = 1$:
\blgn
\frac{V_{Lx−1,1} − V_{Lx,1}}{2R}-\frac{V_{Lx,1}}{R}+\frac{V_{Lx,2} − V_{Lx,1}}{2R}+\frac{V_{Lx-1,2} − V_{Lx,1}}{2R}= 0\,.
\elgn
\item 
\textbf{Right top corner point \ra} $i=L_x$, $j = L_y$:
\blgn
\frac{V_{Lx−1,Ly} − V_{Lx,Ly}}{2R}-\frac{V_{Lx,Ly}}{R}+\frac{V_{Lx,Ly−1} − V_{Lx,Ly}}{2R}= 0\,.
\elgn
\item \textbf{Left non-corner edge point \ra} $i = 1$, $j = 2$ to $L_y − 1$:
\blgn 
\frac{V − V_{1,j}}{R}+\frac{V_{2,j} − V_{1,j}}{2R}+\frac{V_{1,j−1} − V_{1,j}}{2R}+\frac{V_{1,j+1} − V_{1,j}}{2R}+\frac{V_{2,j-1}-V_{1,j}}{2R}= 0\,.
\elgn
\item 
\textbf{Right non-corner edge point \ra} $i = L_x$, $j = 2$ to $L_y − 1$:
\blgn
\frac{V_{Lx−1,j} − V_{Lx,j}}{2R}-\frac{V_{Lx,j}}{R}+\frac{V_{Lx,j−1} − V_{Lx,j}}{2R}+\frac{V_{Lx,j+1} − V_{Lx,j}}{2R}+\frac{V_{Lx-1,j+1} − V_{Lx,j}}{2R}= 0\,.
\elgn
\item 
\textbf{Bottom non-corner edge point \ra} $i = 2$ to $L_x$, $j = 1$:
\blgn
\frac{V_{i−1,1} − V_{i,1}}{2R}+\frac{V_{i+1,1} − V_{i,1}}{2R}+\frac{V_{i,2} − V_{i,1}}{2R}+\frac{V_{i-1,2} − V_{i,1}}{2R}= 0\,.
\elgn
\item \textbf{Top non-corner edge point \ra} $i = 2$ to $L_x$, $j = L_y$:
\blgn
\frac{V_{i−1,Ly} − V_{i,Ly}}{2R}+\frac{V_{i+1,1} − V_{i,Ly}}{2R}+\frac{V_{i,Ly−1} − V_{i,Ly}}{2R}+\frac{V_{i+1,Ly−1} − V_{i,Ly}}{2R}= 0\,.
\elgn
\item \textbf{Non-border inner point \ra} $i = 2$ to $L_x − 1$, $j = 2$ to $L_y− 1$:
\blgn
\frac{V_{i−1,j} − V_{i,j}}{2R}+\frac{V_{i+1,j} − V_{i,j}}{2R}+\frac{V_{i,j−1} − V_{i,j}}{2R}+\frac{V_{i,j+1} − V_{i,j}}{2R}+\frac{V_{i-1,j+1} − V_{i,j}}{2R}+\frac{V_{i+1,j-1} − V_{i,j}}{2R}= 0\,.
\elgn
\end{enumerate}
\subsection{Results:}
\begin{figure}[!htp]
\centering
\subfigure[]{
\includegraphics[height=5cm,clip]{\FIGDIR/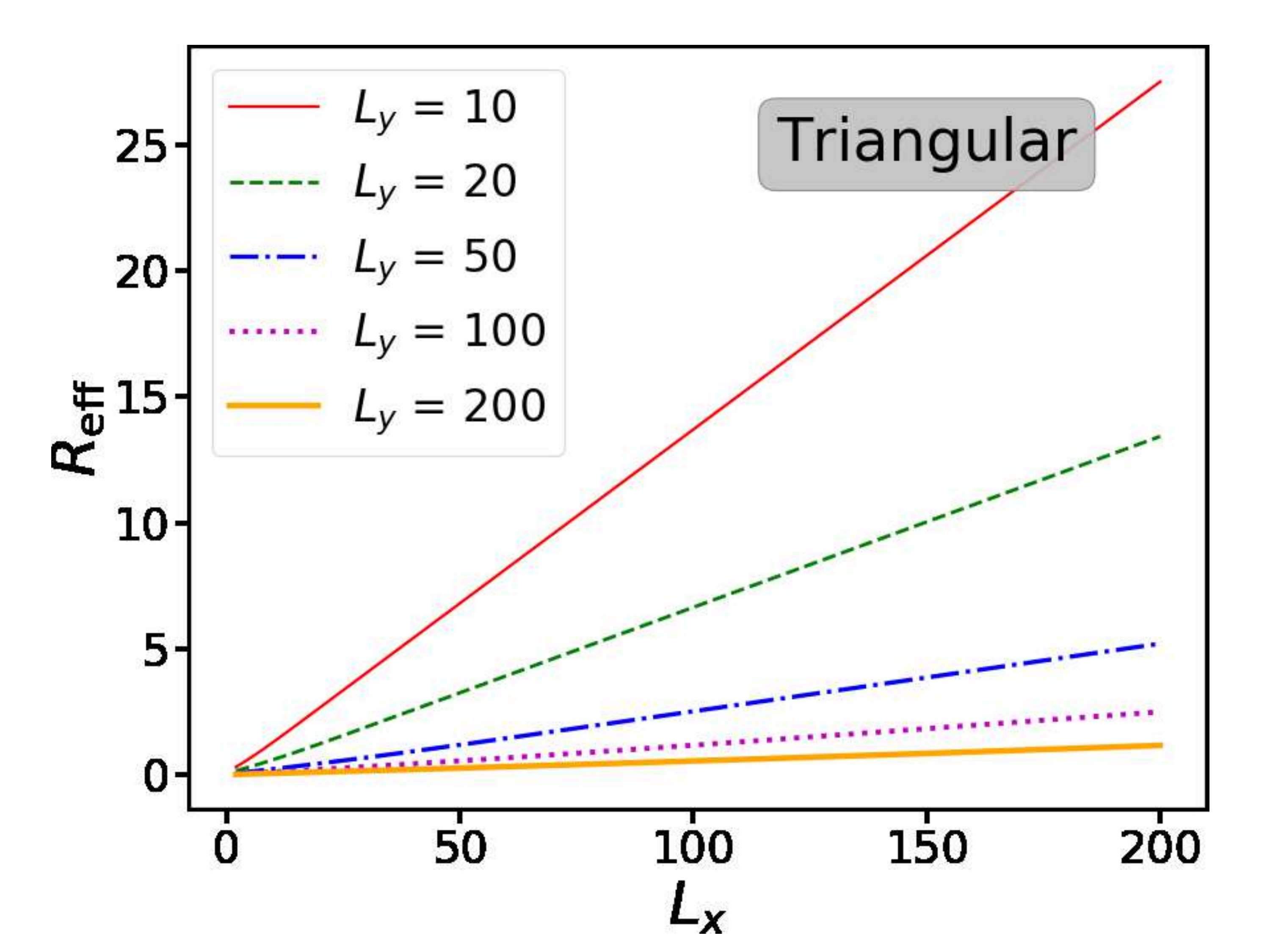}
\label{fig:Reff:vs:Lx:tri}
}
\subfigure[]{
\includegraphics[height=5cm,clip]{\FIGDIR/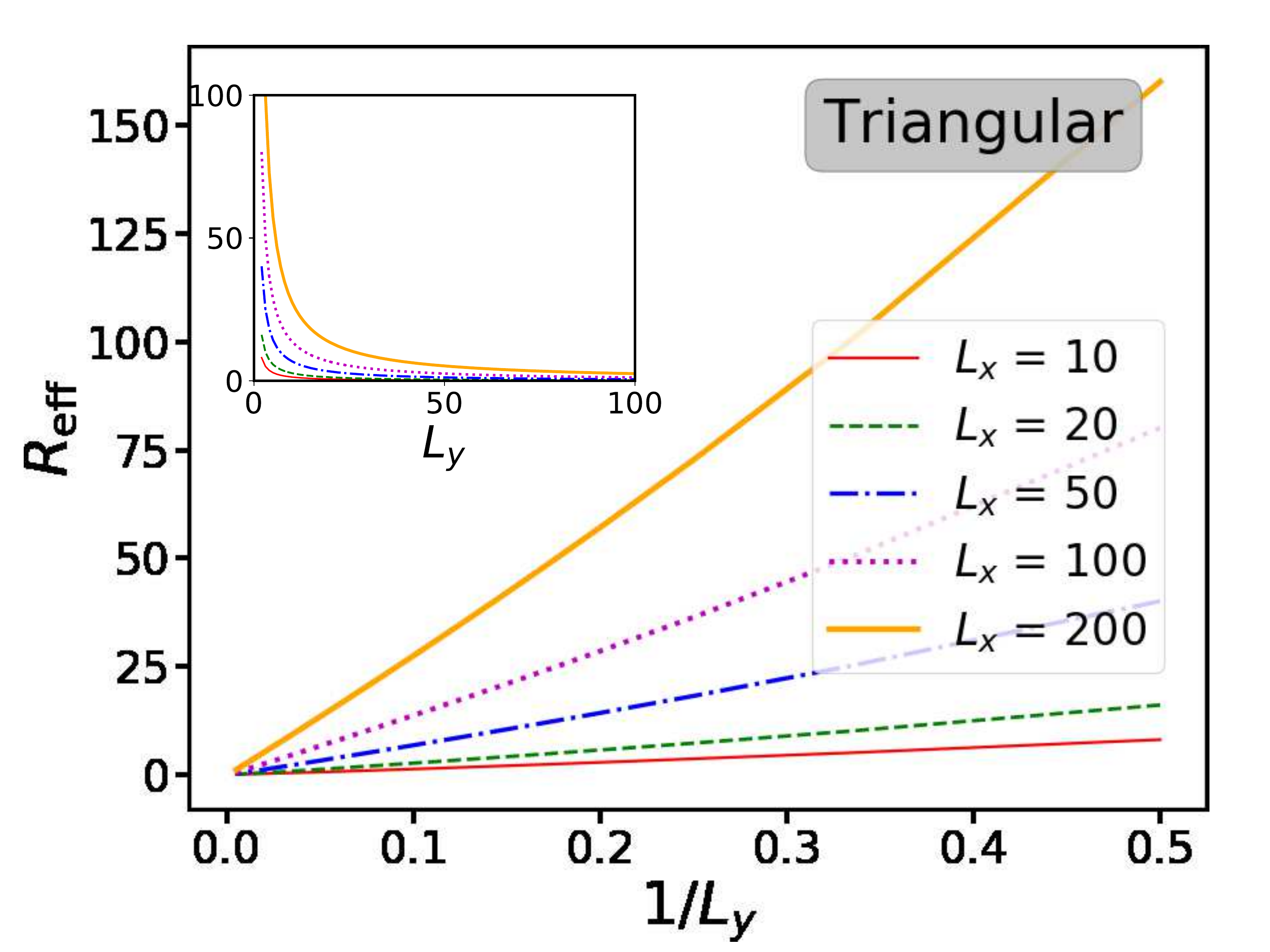}
\label{fig:Reff:vs:Ly:tri}
}
\subfigure[]{
\includegraphics[height=5cm,clip]{\FIGDIR/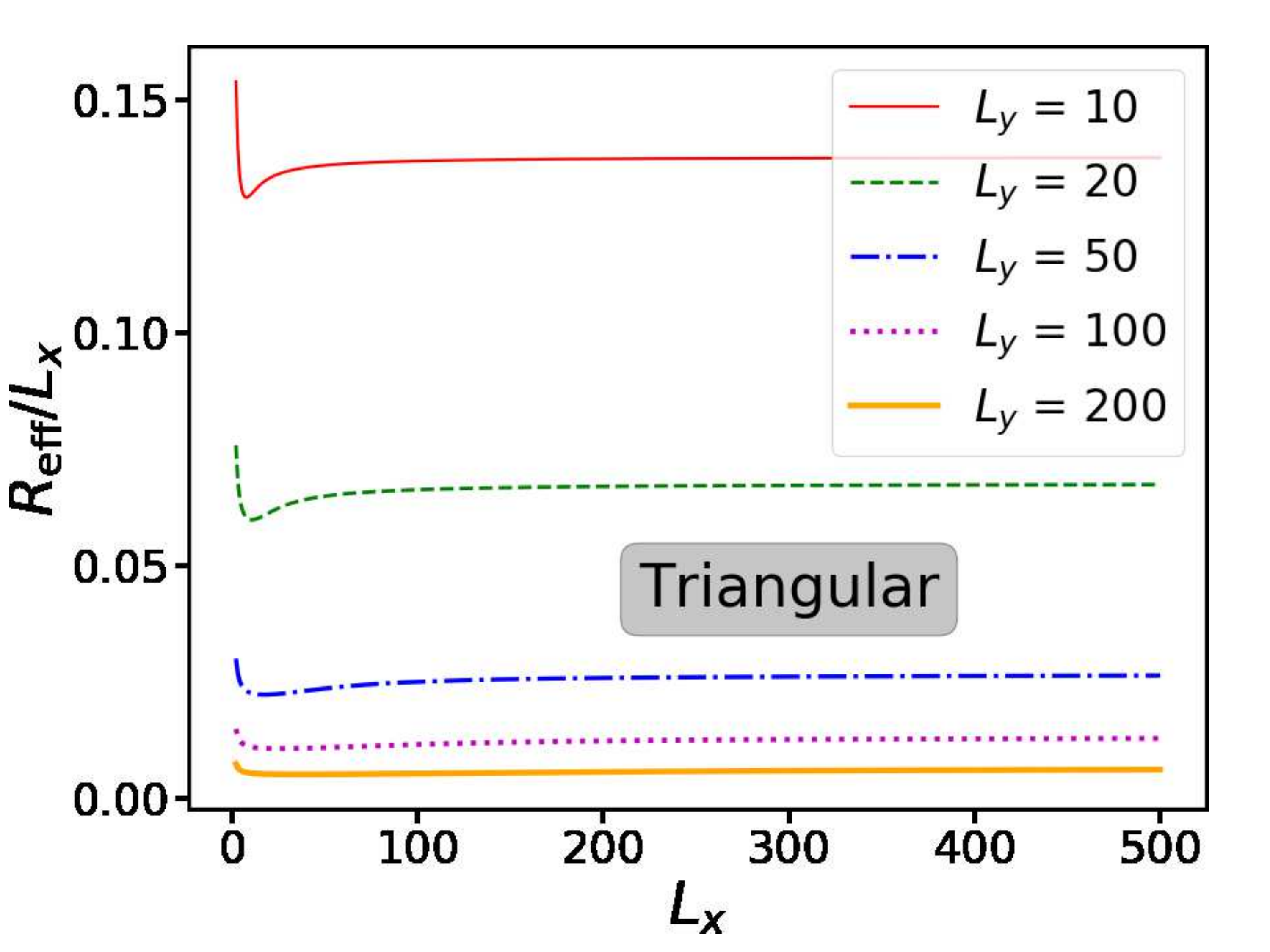}
\label{fig:ReffbyLx:vs:Lx:tri}
}
\subfigure[]{
\includegraphics[height=5cm,clip]{\FIGDIR/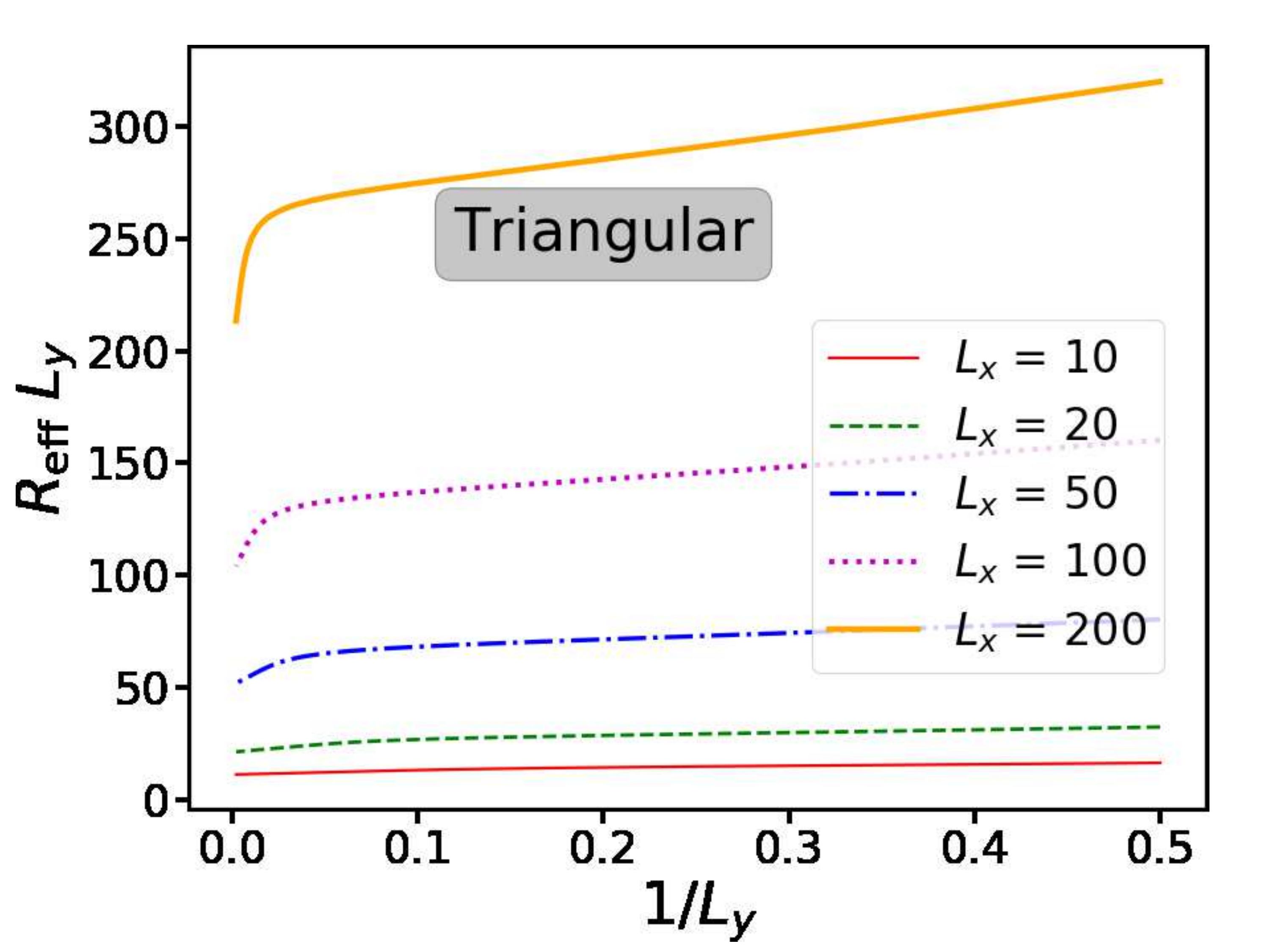}
\label{fig:ReffLy:vs:Lyinv:tri}
}
\caption{Plot of $R_{\eff}$ against (a) $L_x$ for different values of $L_x$ and (b)  $1/L_y$ (inset: against $L_y$) for different values of $L_x$ for a triangular lattice. (c) $R_{\eff}/L_x$ vs $L_x$ plots show that $\Reff$ depends linearly on $L_x$ only at large $L_x$ or $L_y$. (d) $R_{\eff}/L_y$ vs $L_y$ plots show that $\Reff$ does not strictly depends linearly on $1/L_y$.}
\end{figure}

After solving the above equations in the matrix form, we plot $R_{\eff}$ against $L_x$ keeping
$L_y$ fixed at various values. We again notice that $\Reff$ increases with $L_x$ and $1/L_y$ (see  \fref{fig:Reff:vs:Lx:tri}). However, unlike the rectangular or hexagonal lattice network, the dependence of $\Reff$ is not strictly linear, rather it only becomes
linear in $L_x$ at large value of $L_x$ or $L_y$ (see \fref{fig:ReffbyLx:vs:Lx:tri}).

\begin{figure}[!htp]
\centering
\subfigure[]{
\includegraphics[height=5cm,clip]{\FIGDIR/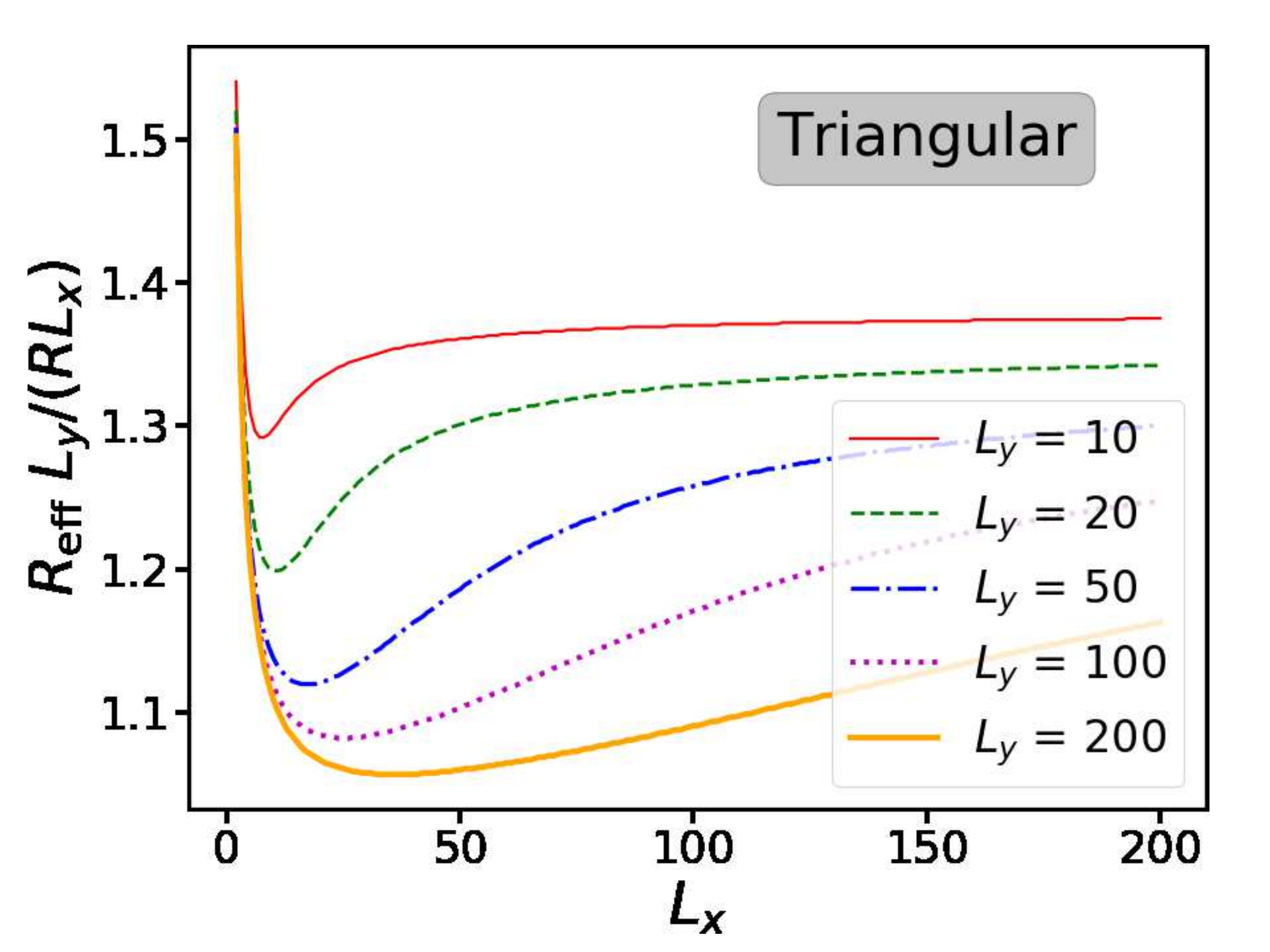}
\label{fig:Rratio:vs:Lx:tri}
}
\subfigure[]{
\includegraphics[height=5cm,clip]{\FIGDIR/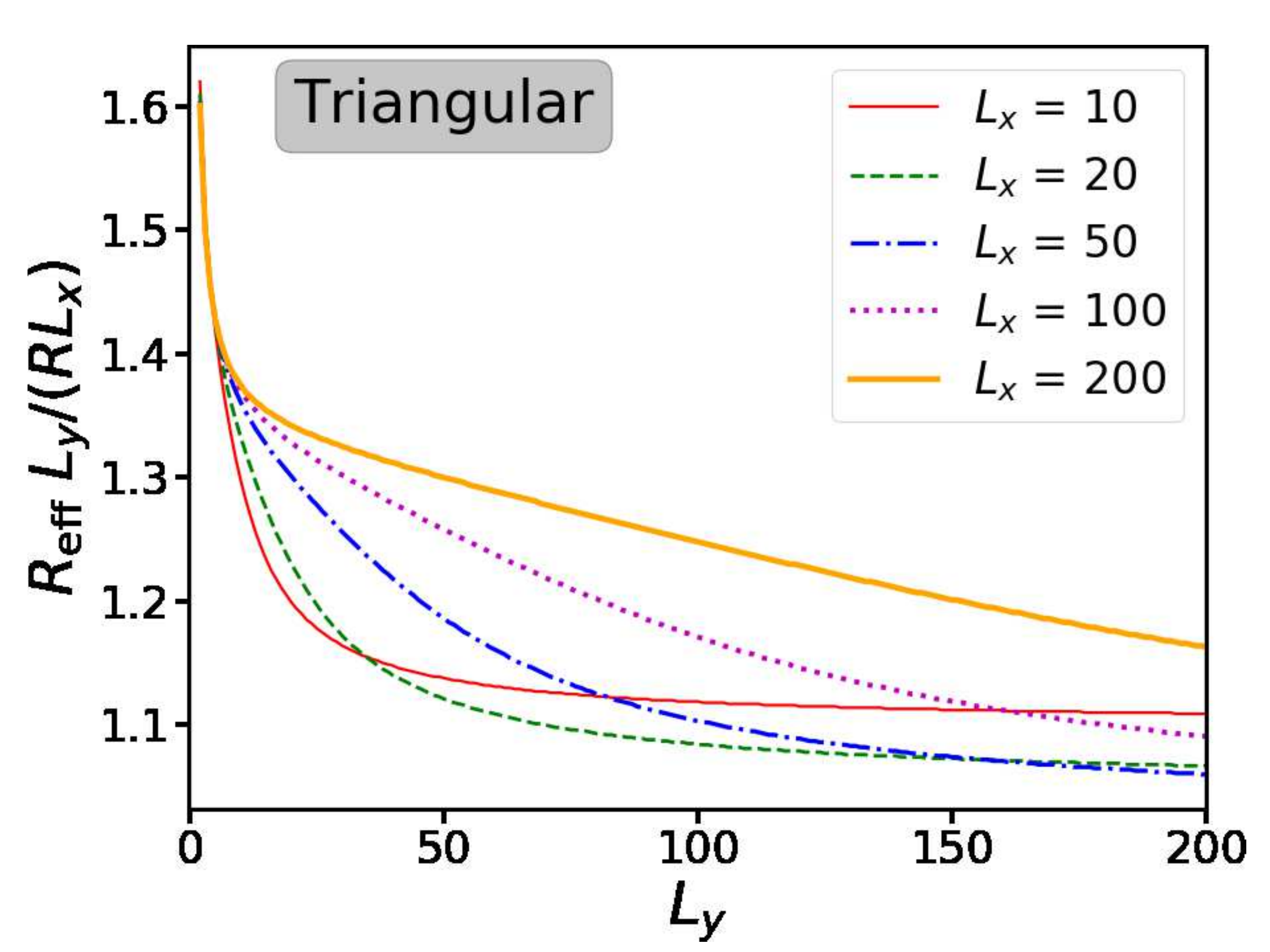}
\label{fig:Rratio:vs:Ly:tri}
}
\caption{Plot of $r=\Reff L_y/(RL_x)$ vs $L_y$ as $L_y$ is varied for different values of $L_x$ for a triangular lattice.}
\label{fig:Rratio:tri}
\end{figure}

Now we look at the ratio $r=\Reff L_y/(R L_x)$ and find that it depends non-trivially on both $L_x$ and $L_y$. As can be noticed from \fref{fig:Rratio:vs:Lx:tri} and \fref{fig:Rratio:vs:Ly:tri}, $r$ approaches a constant value only when $L_x/L_y$  (when $L_x$ varied, $L_y$ fixed) or  $L_y/L_x$  (when $L_y$ varied, $L_x$ fixed) is significantly large (i.e. $L_x$ or $L_y$ approaches the thermodynamic limit compared to the other dimension). However, unlike the earlier two lattice cases, we could not trivially figure out any empirical function or formula for the $\Reff$'s dependence on $L_x$ and $L_y$ for the triangular lattice network. We presume that this non-triviality arises because of the diagonal resistance dependence of the circuit current which is absent in hexagonal and rectangular lattice networks. Also, one should note that hexagonal lattice is a brick-wall lattice, which is a shifted version of a rectangular lattice and hence both lattices bear a similarity in the current flow distribution, and in that sense, triangular lattice network is entirely unique. Only the natural expectation that the effective resistance will be independent of dimension at a large value of $L_x$ and $L_y$ have been reflected in all three different lattice networks discussed in our paper.   

\section{Summary}
Now in \tref{tab:Reff:var:geo}, we briefly summarize $\Reff$'s dependence on the dimensions for 
various network geometries that we discussed already in the previous sections.
\begin{table}[!hbp]
\centering
\begin{tabular}{|c|c|c|c|}
\hline
Network geometry & $L_x$ ($L_y$ fixed) & $L_y$ ($L_x$ fixed) &{Formula}\\
\hline
\hline
Rectangular &{$\propto L_x$} &{$\propto 1/L_y$} &{$R \f{z}{2} L_x/L_y$}\\ 
\hline
Hexagonal (armchair) &{$\propto L_x$} &{$\propto 1/L_y$ at $L_x\ll L_y$} & $R\,\al(z,L_y)\, L_x/L_y$\\
\hline
Triangular &{$\propto L_x$ at $L_x\ll L_y$}  &{not strictly $\propto 1/L_y$ } & $R\,\al(z,L_x,L_y)\, L_x/L_y$\\
\hline
\end{tabular}
\caption{Table for dependence of $\Reff$ on various 2D lattice geometries.}
\label{tab:Reff:var:geo}
\end{table}

Our numerical codes (in Python) are freely available to the public on the Github repository: \url{https://github.com/hbaromega/2D-Resistor-Network}.

%
\section{Experiment: Determining the effective resistance of a resistor network}
\begin{figure}[H]
\centering
\subfigure[]{
\includegraphics[height=4cm,clip]{\FIGDIR/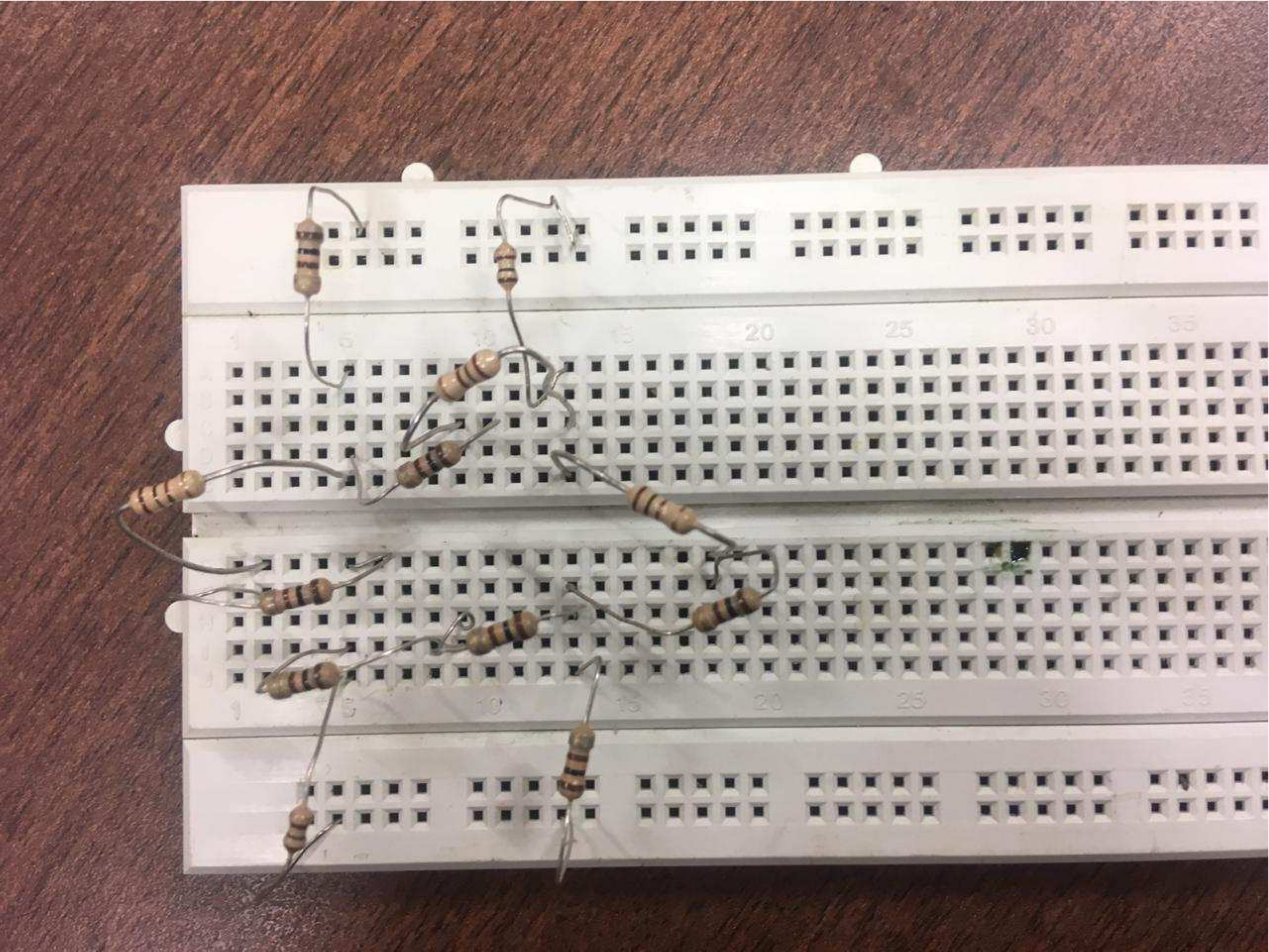}
\label{fig:rect:bb}
}
\subfigure[]{
\includegraphics[height=4cm,clip]{\FIGDIR/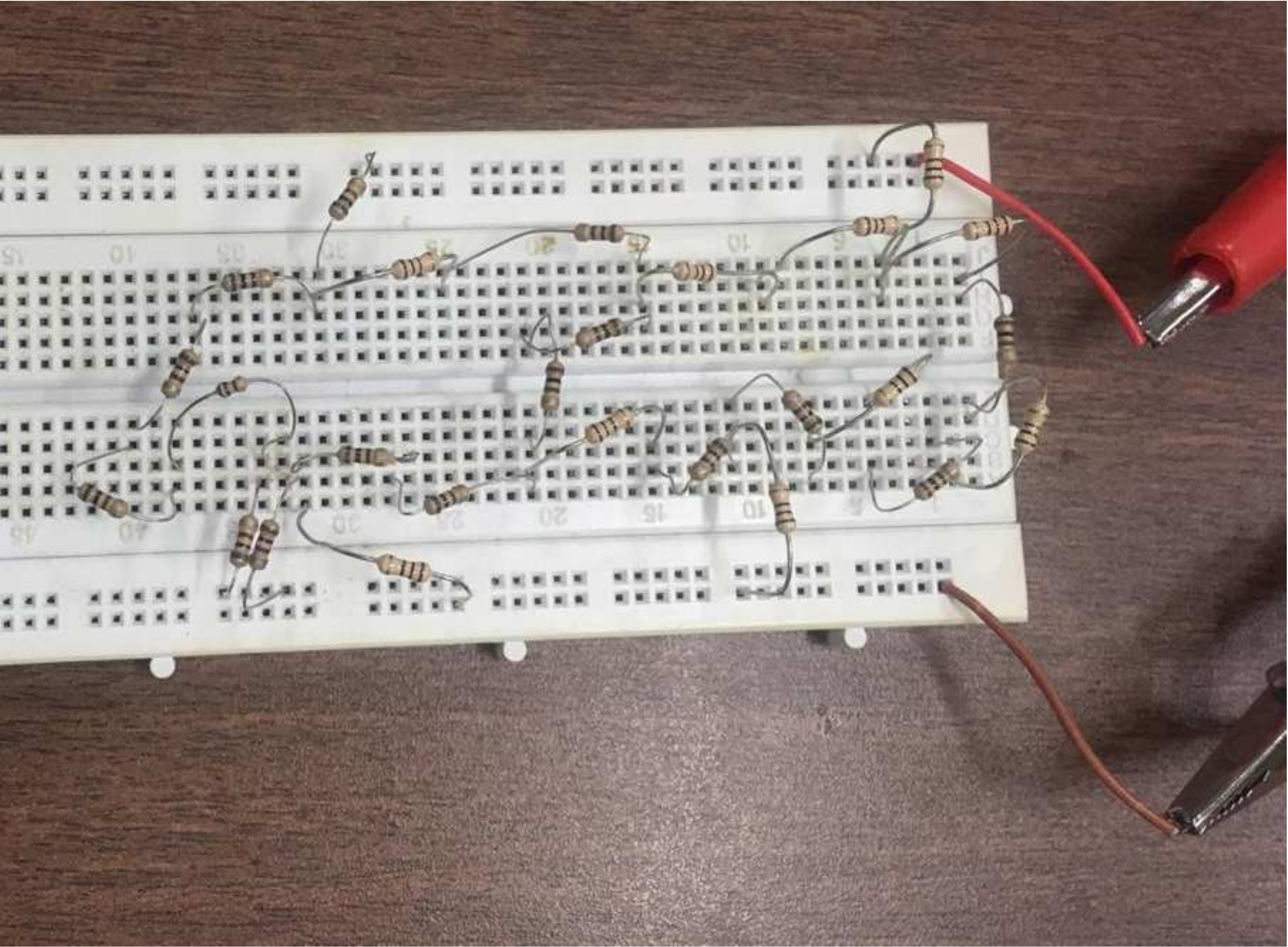}
\label{fig:hex:bb}
}
\\
\subfigure[]{
\includegraphics[height=6cm,clip]{\FIGDIR/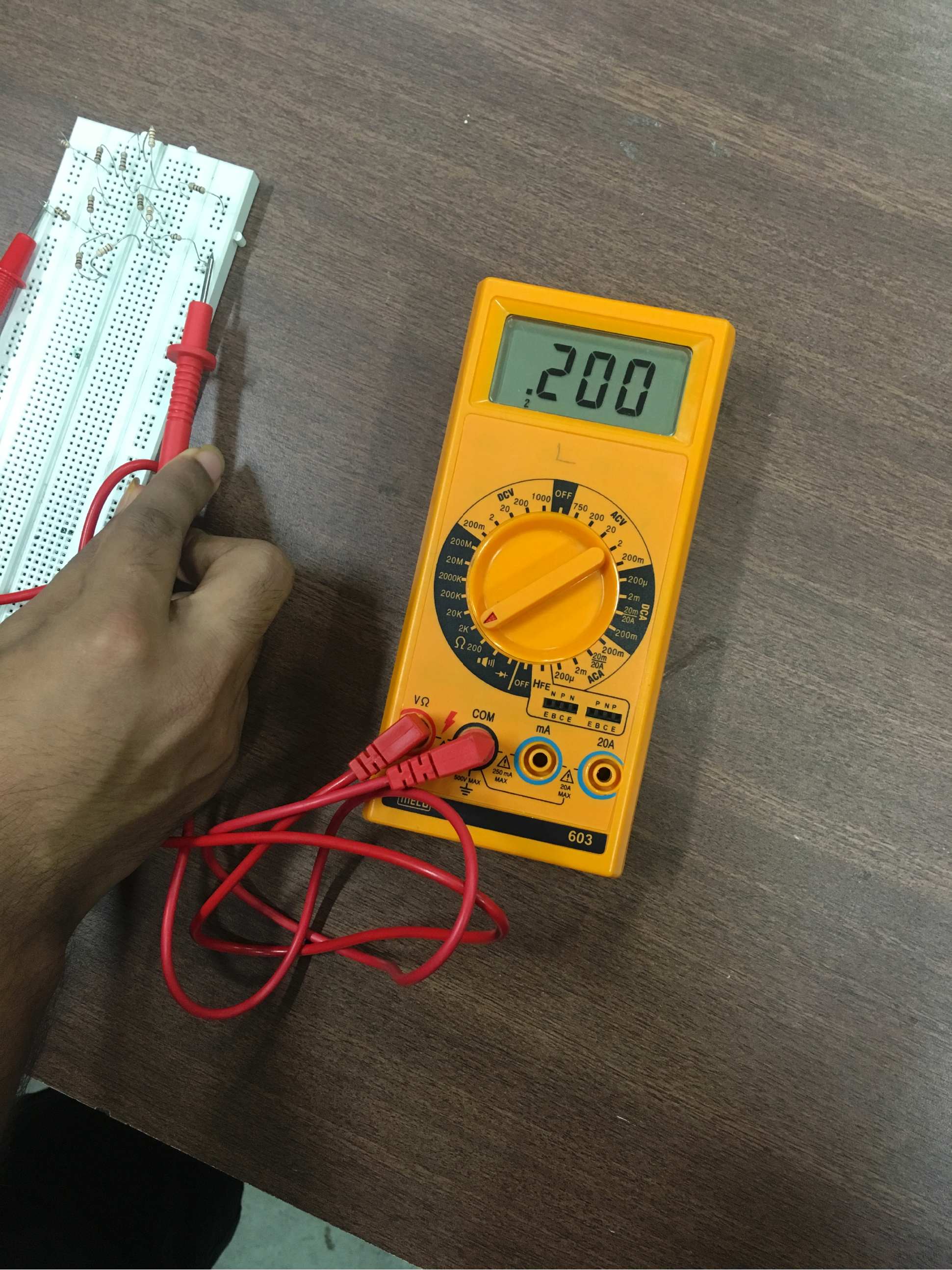}
\label{fig:multimeter_square}
}
\subfigure[]{
\hspace{1cm}\includegraphics[height=6cm,clip]{\FIGDIR/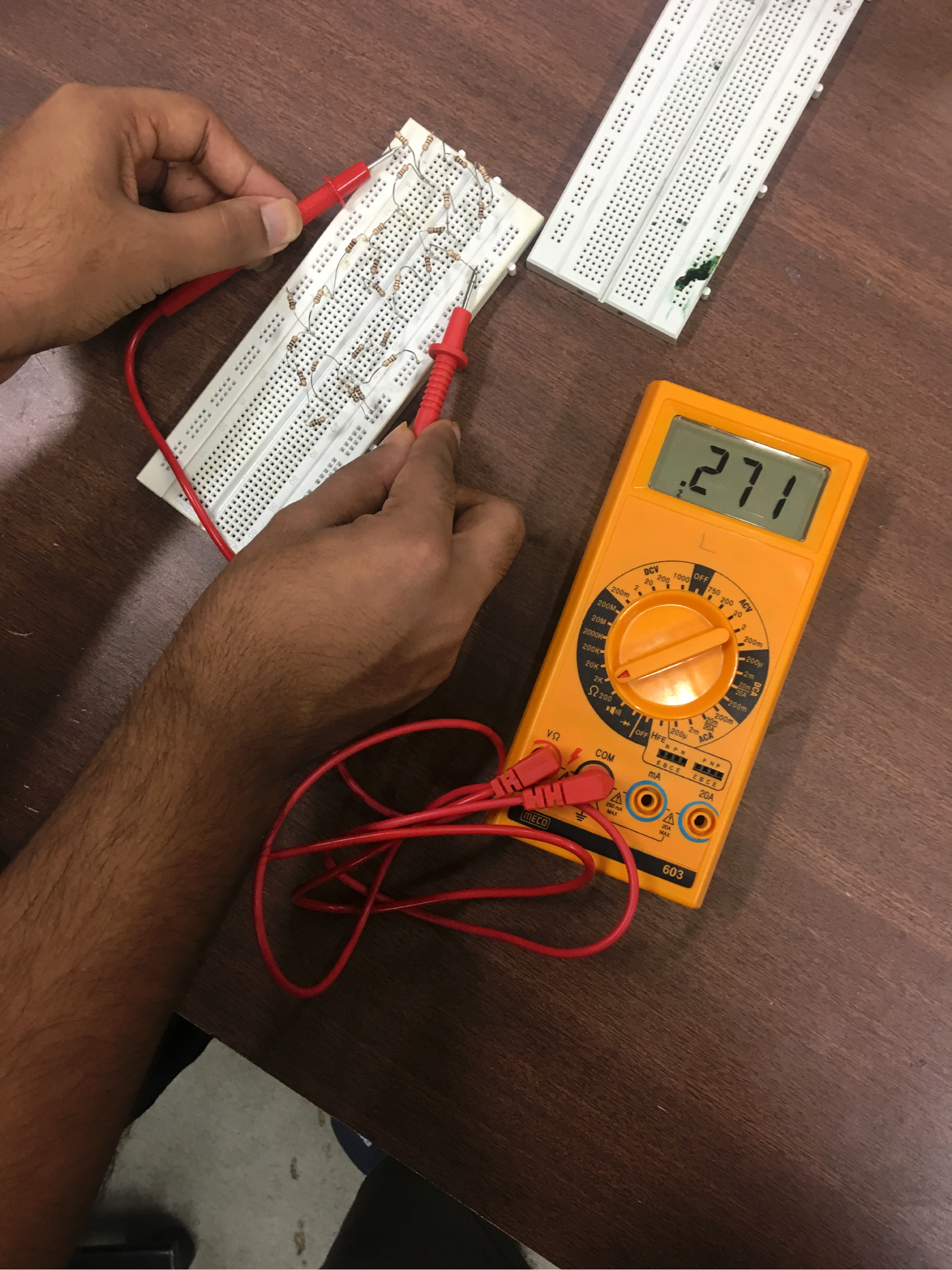}
\label{fig:multimeter_hexagon}
}
\caption{(a) Setup of $2\times2$ square network  and (b) $2\times 2$ hexagonal network on a breadboard. Multimeter's reading of $\Reff$ for  (c) the square and (d) hexagonal networks. Each resistor in the networks has resistance $R=100$ $\Omega$.}
\end{figure}

\begin{figure}[!htp]
\centering
\subfigure[]{
\includegraphics[height=6cm, clip]{\FIGDIR/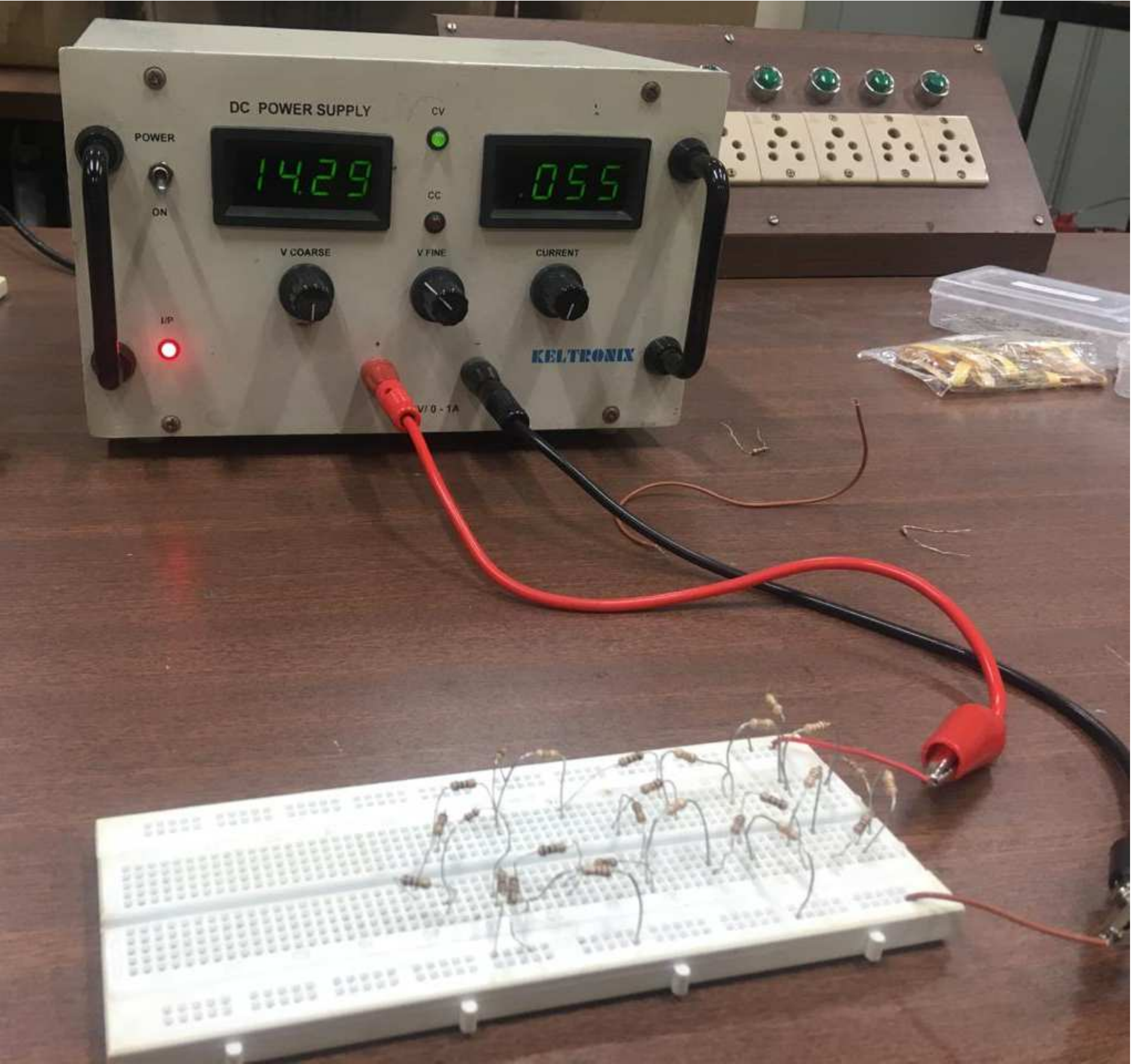}
\label{fig:expt_setup}
}
\subfigure[]{
\hspace{1cm}
\includegraphics[height=6cm,clip]{\FIGDIR/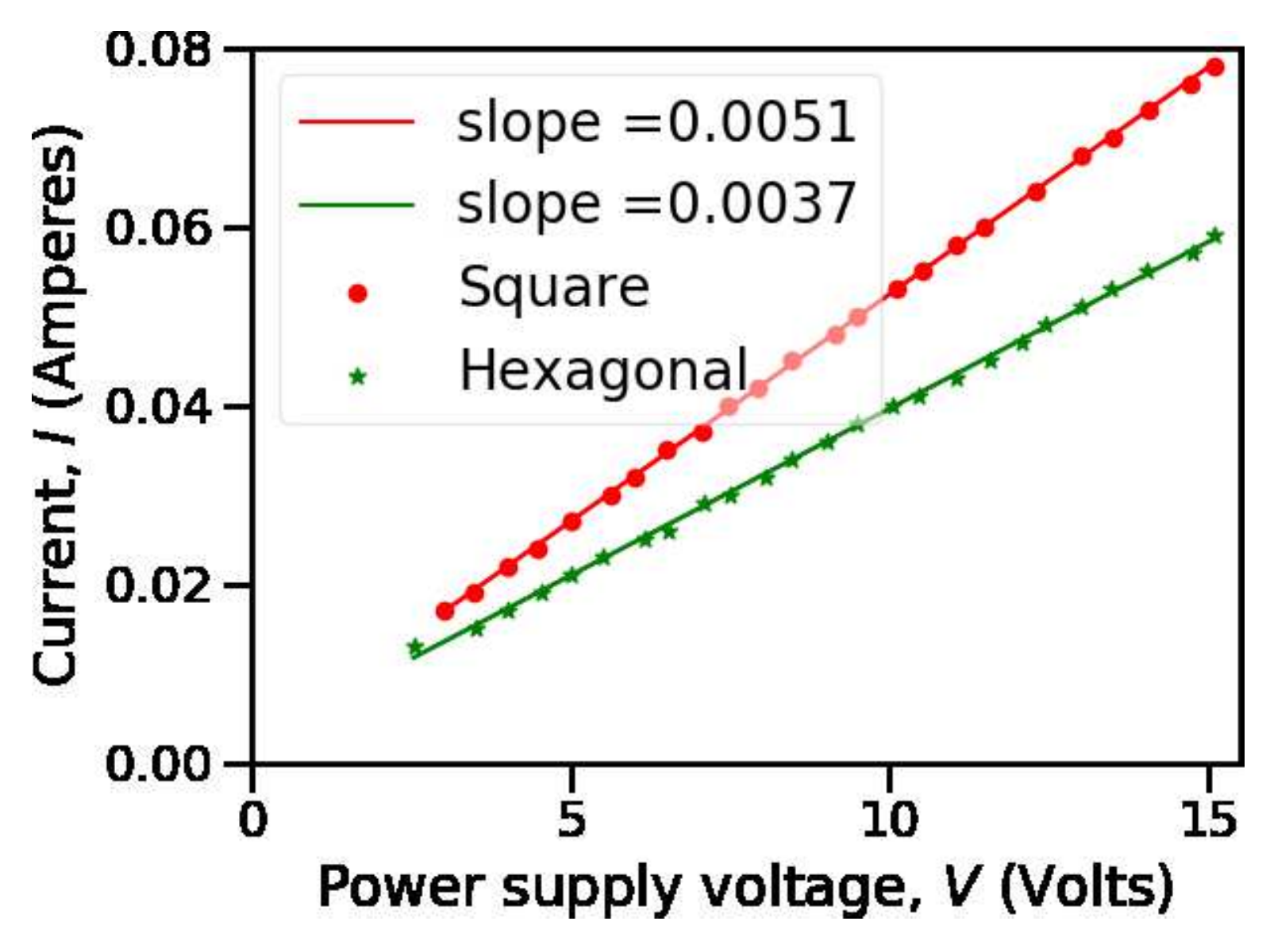}
\label{fig:I:vs:V:w:LR}
}
\caption{(a) Setup for determining internal resistance using voltage regulator. The resistance is obtained by dividing the value of voltage by the value of current. (b) Current vs voltage plots for square and hexagonal lattices. The slopes $s$ of the curves estimate the values of $\Reff$ ($\Reff=1/\Geff$). Here we find $\Reff=196.08\,\Omega$ for the square network and $\Reff=270.27\,\Omega$ for the hexagonal network.}
\end{figure}

Our theoretical findings can be easily verified by setting up simple circuits made up of resistors of equal magnitudes. We first constructed a $2\times2$ rectangular or square network (case A) and a $2\times 2$ armchair hexagonal network (case B) on a breadboard using equal resistors of resistance 100 $\Omega$. We measured the effective resistances of the networks using a multimeter (MECO 603 Digital Multimeter) and compared it with our theoretical results. In cases A and B, $\Reff$ should be $2R=200$ $\Omega$ and $2.7143 R=271.43$ $\Omega$ respectively (see ~\sref{sec:rect:network} and ~\sref{sec:hex:network}). Our multimeter readings show 200 $\Omega$ for case A and 271 $\Omega$ for case B respectively, showing consistent agreement with our theoretical predictions (\fref{fig:multimeter_square} and \fref{fig:multimeter_hexagon}).
%
%
We then connect the networks to a DC power supply (manufactured by Keltronix, India) and determined the effective resistance by measuring the voltage and current across the circuit (figure \ref{fig:expt_setup}).
The following are the readings obtained for the $2\times2$ square and hexagonal networks.
\subsection*{Table 1: Rectangular/square case}
\begin{table}[H]
\centering
\begin{tabular}{|c|c|}
\hline
\textbf{V} (Volts) &  \textbf{I} (Amperes) \\ \hline
3.01   & 0.017	\\ \hline
3.48	& 0.019	\\ \hline
4.01	& 0.022	\\ \hline
4.49	& 0.024	\\ \hline
5.01	& 0.027	\\ \hline
5.63	& 0.03	\\ \hline
6.01	& 0.032	\\ \hline
6.51	& 0.035	\\ \hline
7.06	& 0.037	\\ \hline
7.47	& 0.04	\\ \hline
7.95	& 0.042	\\ \hline
8.47	& 0.045	\\ \hline
9.16	& 0.048	\\ \hline
9.49	& 0.05	\\ \hline
10.13	& 0.053	\\ \hline
10.53	& 0.055	\\ \hline
11.06	& 0.058	\\ \hline
11.49	& 0.06	\\ \hline
12.3	& 0.064	\\ \hline
13.02	& 0.068 \\ \hline
13.5	& 0.07	\\ \hline
14.06	& 0.073	\\ \hline
14.71	& 0.076	\\ \hline
15.1	& 0.078	\\ \hline
\end{tabular}
\end{table}

\subsection*{Table 2: Hexagonal (armchair) case}
\begin{table}[H]
\centering
\begin{tabular}{|c|c|}
\hline
\textbf{V} (Volts) &  \textbf{I} (Amperes)\\ \hline
2.54	& 0.013	\\ \hline
3.52	& 0.015\\ \hline
4	& 0.017	\\ \hline 
4.55	& 0.019\\ \hline
5	& 0.021	\\ \hline
5.52	& 0.023	\\ \hline
6.15	& 0.025\\ \hline
6.52	& 0.026	\\ \hline
7.08	& 0.029	\\ \hline
7.5	& 0.03	\\ \hline
8.06	& 0.032 \\ \hline
8.48	& 0.034	\\ \hline
9.02	& 0.036	\\ \hline
9.5	& 0.038	\\ \hline
10.06	& 0.04	\\ \hline 
10.45	& 0.041 \\ \hline
11.05	& 0.043	\\ \hline
11.59	& 0.045	\\ \hline
12.09	& 0.047	\\ \hline
12.46	& 0.049	\\ \hline
13.02	& 0.051	\\ \hline
13.49	& 0.053	\\ \hline
14.03	& 0.055	\\ \hline
14.74	& 0.057	\\ \hline
15.1	& 0.059	\\ \hline
\end{tabular}
\end{table}
%
%

From the above two tables, we plot the $I$-$V$ (current vs voltage) curves and fit each of them with linear regression lines using the least square method~\cite{book:montgomery:etal15}. The slopes of the regression lines estimate the values of conductance, $\Geff=1/\Reff$. We find  $\Geff=0.0051\,\Omega\inv$ and $\Geff=0.0037\,\Omega\inv$, implying  $\Reff=196.08\,\Omega$ and $\Reff=270.27\,\Omega$ for case $A$ (square) and $B$ (hexagonal) respectively. The values are slightly off the theoretical values: 1.96\% below for case A and 0.43\% below for case B. In case A, the root mean square error (RMSE) and coefficient of determination ($R^2$ score)~\cite{book:montgomery:etal15} of the regression line are  5.76$\times 10^{-8}$ and 0.999831 respectively. The same for case B are 1.6$\times 10^{-7}$ and 0.999139 respectively. The values reflect that regression lines have reasonably high accuracy. The lines, however, show very small finite intercepts of values $0.00158$  and $0.00226$ Amperes. These offset values possibly originate from the resistances in the circuit connection (breadboard and wire connections to the power supply) since we have already checked that the networks accurately produce the theoretical result when measured separately with a multimeter. Thus as inference, we must say that our experiment validates the theory within very low error bars. 
The Python codes of our experimental plots and regression analysis can be found at \url{https://github.com/hbaromega/2D-Resistor-Network/tree/master/EXPT}. 

\section{Outlook}
The detailed but simple derivations of finite size lattice networks of three distinct geometries and the discussed simple experiment on a breadboard setup offer very easy and effective way to teach \emph{network analysis} to students or even adults since the background requirement is minimal (only Kirchhoff's laws and a programming language). Therefore, this can be added to one of the earlier proposed curricula~\cite{porebska:schmidt:zegarmistrz:icses14} in this regard. Later, such studies can be connected to the graph theory since graphs offer visual appeal to one's learning and conceptual comprehensibility.

\section*{Acknowledgments}
HB and RCM owe to the NIUS Camp 2019, HBCSE, Mumbai, which made the project to be worked out and successful. They also thank to Dr. Rajesh Khaprade and Dr. Praveen Pathak for providing the necessary hospitality and experimental facilities.

\appendix
\section{Linearization of V-matrix}
The KCL equations contain two-dimensional $V_{i,j}$ elements. For the rectangular or triangular lattice, when we linearize it to a one-dimensional vector or column matrix, we take either of these two mappings:\\
{\bf Mapping 1:}
\blgn
V_{1,1},\cdots,V_{L_x,1} &\to V_1,\cdots,V_{L_x}\,.\non\\
V_{1,2},\cdots,V_{L_x,2} &\to V_{L_x+1},\cdots,V_{2L_x}\,.\non\\
&\vdots\non\\
V_{1,L_y},\cdots,V_{L_x,L_y}&\to V_{(L_y-1)L_x+1},\cdots,V_{L_y L_x}\,.\non\\
\text{Generically,}\quad (i,j) &\to (j-1)L_x+i\,.
\elgn 
{\bf Mapping 2:} 
\blgn
V_{1,1},\cdots,V_{1,L_y} &\to V_1,\cdots,V_{L_y}\,.\non\\
V_{2,1},\cdots,V_{2,L_y} &\to V_{L_y+1},\cdots,V_{2L_y}\,.\non\\
\vdots\non\\
V_{L_x,1},\cdots,V_{L_x,L_y}&\to V_{(L_x-1)L_y+1},\cdots,V_{L_x L_y}\,.\non\\
\text{Generically,}\quad (i,j) &\to (i-1)L_y+j\,.
\elgn 

Now a typical equation such as \eref{eq:KCL:rect:rearranged:1st} looks like
\blgn
\al V_1 + \bt V_2 +\cdots+\g V_{L_x} = I_1\,. 
\elgn 
which can be recast as
\blgn
G_{11} V_1 + G_{12} V_2 + \cdots + G_{1L_x} V_{L_x}=I_1\,. 
\elgn
Generically this can be written as
\blgn
G_{i1} V_1 + G_{i2} V_2 + \cdots + G_{iL_x} V_{L_x}=I_i\,.
\label{eq:KCL:gen}
\elgn
which builds the matrix form:
\blgn
\begin{bmatrix}
G_{11} &G_{12} &\cdots &G_{1 N_L} \\
G_{21} &G_{22} &\cdots &G_{1 N_L} \\
\vdots &\vdots &\ddots &\vdots\\
G_{N_L 1}  &G_{N_L 2} &\cdots &G_{N_L N_L}
\end{bmatrix}
\begin{bmatrix}
V_1 \\
V_2 \\
\vdots\\
V_{N_L}
\end{bmatrix}
=
\begin{bmatrix}
I_1 \\
I_2 \\
\vdots\\
I_{N_L}
\end{bmatrix}\,.
\elgn
where $N_L\equiv L_x L_y$.

\subsection*{Finding corresponding row and column of ${\bf G}$, given row of ${\bf I}$:}
Since each lattice point follows a particular KCL depending on its neighborhood and voltage connection,
the rank of that lattice (reflected by the row or index $i$ of current vector $I$ in \eref{eq:KCL:gen}) in the mapped 1D array will denote the row of  ${\bf G}$ and the index of ${\bf V}$ (which is a vector or column matrix) will yield the column of ${\bf G}$.  

\subsection*{Mapping in the hexagonal lattice case:}
Since we introduce another index $k$ in the armchair hexagonal lattice, we extend the linear mapping as 
\blgn
(i,j,k) \to i+(j-1)L_x + kL_x L_y\,.
\elgn

One can check the mapping conserves the total number of points $N_\site=L_x (2L_y + 1)$:

$k=0$ case:\\
\blgn
V_{1,1,0},\cdots,V_{L_x,1,0} &\to V_1,\cdots,V_{L_x}\,.\non\\
V_{1,2,0},\cdots,V_{L_x,2,0} &\to V_{L_x+1},\cdots,V_{2L_x}\,.\non\\
&\vdots\non\\
V_{1,L_y,0},\cdots,V_{L_x,L_y,0}&\to V_{(L_y-1)L_x+1},\cdots,V_{L_y L_x}\,.\non\\
\elgn
$k=1$ case:\\
\blgn
V_{1,1,1},\cdots,V_{L_x,1,1} &\to V_{L_xL_y+1},\cdots,V_{(L_y+1)L_x}\,.\non\\
V_{1,2,1},\cdots,V_{L_x,2,1} &\to V_{(L_y+1) L_x+1},\cdots,V_{(L_y+2)L_x}\,.\non\\
&\vdots\non\\
V_{1,L_y,1},\cdots,V_{L_x,L_y,1}&\to V_{(2L_y-1)L_x+1},\cdots,V_{2L_y L_x}\,.\non\\
V_{1,L_y+1,1},\cdots,V_{L_x,L_y+1,1}&\to V_{(2L_y L_x+1},\cdots,V_{(2L_y+1) L_x}\,.\non\\
\elgn

\section*{References}
\bibliographystyle{unsrt}
\bibliography{refs_rn2d}
%
\end{document}